\documentclass[twocolumn,showpacs,superscriptaddress]{revtex4}

\usepackage{amsmath}
\usepackage{amssymb}
\usepackage{epsfig}

\newcommand{\GEANT}{{\tt GEANT4}}
\newcommand{\FLUKA}{{\tt FLUKA}}
\newcommand{\PEANUT}{{\tt PEANUT}}

\begin{document}

\title{Nucleon Decay Searches with large Liquid Argon TPC
Detectors at Shallow Depths: atmospheric neutrinos
and cosmogenic backgrounds}

\newcommand{\granada}{ Dpto de F{\'\i}sica Te\'orica y del Cosmos \& C.A.F.P.E.,
Universidad de Granada, Granada, Spain} 
\newcommand{\ETH}{Institute for Particle Physics, ETH H\"onggerberg, Z\"urich, Switzerland}

\affiliation{\granada}
\affiliation{\ETH}

\author{A.~Bueno}
\affiliation{\granada}
\author{Z.~Dai}
\affiliation{\ETH}
\author{Y.~Ge}
\affiliation{\ETH}
\author{M.~Laffranchi}
\affiliation{\ETH}
\author{A.J.~Melgarejo}
\affiliation{\granada}
\author{A.~Meregaglia}
\affiliation{\ETH}
\author{S.~Navas}
\affiliation{\granada}
\author{A.~Rubbia}
\affiliation{\ETH}

\begin{abstract}
Grand Unification of the strong, weak and electromagnetic interactions
into a single unified gauge group is an extremely appealing idea which has
been vigorously pursued theoretically and experimentally for many years.
The detection of proton or bound-neutron decays would represent its
most direct experimental evidence.
In this context, we studied the physics potentialities
of very large underground Liquid Argon Time Projection Chambers (LAr TPC).
We carried out a detailed simulation of signal efficiency and
background sources, including atmospheric neutrinos and cosmogenic
backgrounds. We point out that a liquid Argon TPC,
offering good granularity and energy resolution, low particle detection threshold,
and excellent background discrimination, should  
yield  very good signal over background ratios in many possible
decay modes, allowing to reach partial lifetime sensitivities in
the range of $10^{34}-10^{35}$~years with exposures up to 1000~kton$\times$year,
 often in quasi-background-free conditions optimal for discoveries
 at the few events level, corresponding
to atmospheric neutrino background rejections of the order of $10^5$.
Multi-prong decay modes like e.g. $p\rightarrow \mu^- \pi^+ K^+$
or $p\rightarrow e^+\pi^+\pi^-$ and channels involving kaons like
e.g. $p\rightarrow K^+\bar\nu$, $p\rightarrow e^+K^0$ and $p\rightarrow \mu^+K^0$
are particularly suitable, since liquid
Argon imaging
provides typically an order of magnitude improvement in efficiencies for similar
or better background conditions compared to Water Cerenkov detectors.
Up to a factor 2 improvement in efficiency is expected for modes like $p\rightarrow e^+\gamma$
and $p\rightarrow \mu^+\gamma$ thanks to the clean photon identification
and separation from $\pi^0$. Channels like $p\rightarrow e^+\pi^0$ or $p\rightarrow \mu^+\pi^0$,
dominated by intrinsic nuclear effects,
yield similar efficiencies and backgrounds as in Water Cerenkov detectors. 
Thanks to the self-shielding and 3D-imaging properties of the liquid Argon TPC,
this result remains valid even at shallow depths where
cosmogenic background sources are important.
We consider the possibility of a very large area annular active muon veto shield in order to
further suppress cosmogenic backgrounds at shallow depths. In conclusion, we find that
this class of detectors does not necessarily require very deep underground laboratories, like those typically
encountered in existing or planned sites, to perform very sensitive nucleon decay searches. 
In addition to a successful completion of the required R\&D necessary to reach 
a relevant liquid Argon mass scale of 100~kton in a cost-effective way, we finally stress  the
importance of an experimental verification of the liquid Argon TPC physics potentialities
to detect, reconstruct and classify events in the relevant GeV~energy range. 
This experimental verification will require in addition
to possible specific tests in charged particle beams, the
collection of neutrino event samples with statistics in the range of 100'000 events
or more, accessible e.g. with a medium-sized detector at near sites of long
baseline artificial neutrino beams.
\end{abstract}

\date{\today}
\maketitle

\section{Introduction}
\label{sec:intro}

Grand Unification (GU) of the strong, weak and electromagnetic interactions
into a single unified gauge~\cite{gut,Georgi:1974sy} is an extremely appealing solution which has
been vigorously pursued, theoretically and experimentally, for many years.
An experimental hint in its favor is the apparent merging of the three
coupling constants at a large
energy scale ($\sim10^{16}$ GeV) when low energy measurements are
extrapolated~\cite{Amaldi:1991cn}.
On the other hand, the most convincing experimental evidence for GU would
be the direct observation of baryon number violation~\cite{Langacker:1980js}. 
The experimental search for 
decays of protons or bound-neutrons
is therefore one of the most important and unsolved
problem of particle physics. 

In the simplest Grand Unified Theories (GUT), nucleon decay proceeds via 
an exchange of a massive boson $X$ between
two quarks in a proton or in a bound neutron.  In this reaction, one quark
transforms into a lepton and another into an anti-quark which binds with a
spectator quark creating a meson.  
According to the experimental results from
Super-Kamiokande~\cite{Shiozawa:1998si,Shiozawa:2003it} constraining
the partial decay to 
$\tau/B(p \rightarrow e^+ \pi^0) > 5.4 \times 10^{33}$~years (90\%C.L.), the minimal
SU(5)~\cite{Georgi:1974sy}, predicting a proton
lifetime proportional to $\alpha^{-2}M_X^4$ where $\alpha$
is the unified coupling constant and $M_X$ the mass of
the gauge boson $X$, seems definitely ruled out.

Supersymmetry, motivated by the so-called ``hierarchy problem'',
 postulates that for every SM particle, there is a
corresponding ``superpartner'' with spin differing by 1/2 unit from the SM
particle \cite{susy}. In this case,
the unification scale turns out higher, and pushes up the proton lifetime in the $p \rightarrow e^+ \pi^0$
channel up to $10^{36\pm1}$~years, compatible with experimental results.
At the same time, alternative decay channels open up via
dimension-five operator interactions with the exchange of heavy
supersymmetric particles.
In these models, transitions from one quark family in the initial state to the same
family in the final state are suppressed.  
Since the only second or third generation quark
which is kinematically allowed is the strange quark, an anti-strange quark
typically appears in the final state for these interactions. The anti-strange
quark binds with a spectator quark to form a $K$ meson in the final state~\cite{Pati}.
The searches
for decays $p \to \bar{\nu} K^+$, $n \to \bar{\nu} K^0$, $p \to \mu^+ K^0$ and $p \to e^+ K^0$ 
modes were also performed in Super-Kamiokande~\cite{Hayato:1999az,Kobayashi:2005pe}
yielding counts compatible with background expectations, leading to limits on possible
minimal SUSY SU(5) models~\cite{Dimopoulos:1981zb,Sakai:1982pk,Hisano:1993jj}.
The theoretical predictions, however, vary
widely, since there are many new unknown parameters
introduced in these models.

Other alternative models have been discussed in the litterature~\cite{Nath:1985ub, Nath:1998kg, Shafi:1999vm, Lucas:1997bc,
Pati:2003qi, Babu:1998js, Babu:1998wi, Pati:2000wu,Ellis:2002vk,Arkani-Hamed:2004yi, Hebecker:2002rc,
Alciati:2005ur, Klebanov:2003my} (see Table~\ref{tab1}). 
In addition to the above mentioned GUTs, other supersymmetric SUSY-GUT, SUGRA unified models, unification based on extra dimensions, and string-M-theory models
 are also possible (see Ref.~\cite{Nath:2006ut} for a recent review).  
 All these models predict nucleon instability at some level.
 Finally, it is also worth noting that 
 theories without low-energy super-symmetry~\cite{Arkani-Hamed:2004yi,Wiesenfeldt:2006ut}
predict nucleon decay lifetimes in the range $10^{35\pm1}$~years.
 
 \begin{table*}
\centering
{\small
\begin{tabular}{|l|l|l|l|}\hline
Model & Ref. & Modes & $\tau_N$ (years)\\ \hline \hline
Minimal $SU(5)$ & Georgi, Glashow \cite{Georgi:1974sy} & $p\to e^+ \pi^0$ & $10^{30}-10^{31}$\\ \hline
Minimal SUSY $SU(5)$ & Dimopoulos, Georgi \cite{Dimopoulos:1981zb}, Sakai \cite{Sakai:1982pk} &  $p\to \bar{\nu}K^+$ & \\ 
&Lifetime Calculations: Hisano, &$n\to \bar{\nu}K^0$ & $10^{28}-10^{32}$\\
& Murayama, Yanagida \cite{Hisano:1993jj}&&\\ \hline
SUGRA $SU(5)$ & Nath, Arnowitt \cite{Nath:1985ub, Nath:1998kg} & $p\to \bar{\nu}K^+$ &  $10^{32}-10^{34}$\\ \hline
SUSY $SO(10)$ & Shafi, Tavartkiladze \cite{Shafi:1999vm}& $p\to \bar{\nu}K^+$ & \\
with anomalous& &$n\to \bar{\nu}K^0$ & $10^{32}-10^{35}$\\
flavor $U(1)$ & &$p\to \mu^+K^0$ &\\ \hline
SUSY $SO(10)$ &Lucas, Raby \cite{Lucas:1997bc}, Pati~\cite{Pati:2003qi}&  $p\to \bar{\nu}K^+$ &  $10^{33}-10^{34}$\\ \cline{3-4}
MSSM (std. $d=5$)&&$n\to \bar{\nu}K^0$ &  $10^{32}-10^{33}$\\ \hline
SUSY $SO(10)$&Pati~\cite{Pati:2003qi}&$p\to \bar{\nu}K^+$&$10^{33}-10^{34}$\\
ESSM (std. $d=5$)&&&$\lesssim 10^{35}$\\ \hline
SUSY $SO(10)/G(224)$ & Babu, Pati, Wilczek \cite{Babu:1998js, Babu:1998wi, Pati:2000wu},&  $p\to \bar{\nu}K^+$& $\lesssim 2\cdot 10^{34}$\\\cline{3-4}
MSSM or ESSM  &Pati~\cite{Pati:2003qi}& \multicolumn{2}{l|}{$p\to \mu^+K^0$}Ê\\
(new $d=5$)&&\multicolumn{2}{r|}{$B\sim(1-50)\%$}\\ \hline
SUSY $SU(5)$ or $SO(10)$&Pati~\cite{Pati:2003qi}&$p\to e^+ \pi^0$&$\sim 10^{34.9\pm 1}$\\
MSSM ($d=6$)&&&\\ \hline
Flipped $SU(5)$ in CMSSM& Ellis, Nanopoulos and Wlaker\cite{Ellis:2002vk} & $p\to e/\mu^+\pi^0$ & $10^{35}-10^{36}$\\ \hline
Split $SU(5)$ SUSY & Arkani-Hamed, \emph{et. al.}~\cite{Arkani-Hamed:2004yi} & $p\to e^+\pi^0$ & $10^{35}-10^{37}$\\ \hline
$SU(5)$ in 5 dimensions & Hebecker, March-Russell\cite{Hebecker:2002rc} & $p\to \mu^+K^0$ & $10^{34}-10^{35}$\\
&&$p\to e^+ \pi^0$ &\\ \hline
$SU(5)$ in 5 dimensions & Alciati \emph{et.al.}\cite{Alciati:2005ur} & $p\to \bar{\nu}K^+$ & $10^{36}-10^{39}$\\
option II&& &\\ \hline
GUT-like models from & Klebanov, Witten\cite{Klebanov:2003my}& $p\to e^+ \pi^0$ & $\sim10^{36}$\\
Type IIA string with D6-branes &  & &\\ \hline
\end{tabular}
}
\caption{Summary of the expected nucleon lifetime in different theoretical models.}
\label{tab1}
\end{table*}

Some experimental aspects of
nucleon decay detection were discussed in Ref.~\cite{Rubbia:2004yq}.
Nucleon decay signals are characterized by (a) their topology and
(b) their kinematics. The presence of a lepton
(electron, muon or neutrino) in the final state is expected, and in general few
other particles (two body decays are kinematically favored), and
no other energetic nucleon. The total energy of the event should be
close to the nucleon mass and the total momentum should be 
balanced, with the exception of the smearing introduced by Fermi motion and 
other nuclear effects (nuclear potential, re-scattering, absorption, etc.) for bound
decaying nucleons.

 The search for nucleon decay therefore requires
(1) excellent tracking and calorimetric resolutions to constrain
the final state kinematics and suppress atmospheric neutrino backgrounds, 
(2) particle identification (in particular kaon tagging) for branching mode
identification
(3) very massive detectors and (4) underground locations to
shield against cosmic-ray induced backgrounds, although the exact
required rock overburden depends on the chosen detection technology.
Fine tracking in the low momentum range ($\sim$100--1000~MeV/c)
is fundamental for ($dE/dx$) measurement, particle identification and vertex
reconstruction. 

In order to significantly improve current experimental results,
next generation massive underground detectors satisfying the
above requirements have to be 
considered. Given the variety of predicted decay modes open by the new theories,
the ideal detectors should be as versatile as possible, very good in background
rejection, and at the same time have the largest possible mass. The relevant factor
is in fact M$\times \epsilon$, where M is the detector mass and
$\epsilon$
the signal detection efficiency after cuts to suppress backgrounds, 
which depends on the considered decay
channel.
Hence large masses must be coupled to fine tracking and
excellent calorimetry, to suppress atmospheric neutrino and cosmogenic
backgrounds with a good signal selection efficiency.
Furthermore, the detector should be sensitive to several different channels in order
to better understand the nucleon decay mechanism.
Since there are about $6 \times 10^{32}$ nucleons per kton of mass, the proton
lifetime limit (90\% CL) in case of absence of signal and backgrounds is about
$\tau_p/\text{B} >$ M (kton) $\times$ $\epsilon$  $\times$ T
$\times 10^{32}$ years, where T is the exposure in years and B the assumed
branching fraction for the searched mode. 
Therefore, the required effective mass M$\times \epsilon$ to reach $10^{35}$~years
is in the range of 100~kton assuming T$=$10~years.

Such massive underground detectors will be sort of observatories
for rare physics phenomena like astrophysical neutrino detection and nucleon decay
searches, with possible synergies with existing or new artificial neutrino
beams for improved understanding of neutrino flavor oscillations, including
the possible identification of direct CP-violation in the leptonic sector (see e.g.
Ref.~\cite{Meregaglia:2006du} and references therein).

Among the various options and technologies currently thought of (see Ref.~\cite{nufact06} and
Refs.~\cite{Nakamura:2000tp,Suzuki:2001rb,Jung:1999jq,Diwan:2000yb,deBellefon:2006vq,MarrodanUndagoitia:2006rf,MarrodanUndagoitia:2006qn}),
the Liquid Argon Time Projection 
Chamber~\cite{intro1,Aprile:1985xz,3tons,Cennini:ha,Arneodo:2006ug,t600paper,Amoruso:2004ti,Amoruso:2004dy,Badertscher:2004py,Badertscher:2005te} (LAr TPC) is
a powerful detector for uniform and high accuracy imaging of massive active volumes. 
It is based on the fact that in highly pure Argon, ionization tracks can be drifted
over distances of the order of meters. 
Imaging is provided by position-segmented electrodes at the end of the drift path, continuously recording the 
signals induced. $T_0$ is provided by the prompt scintillation light. 

Early work on the detection of nucleon decays
in liquid Argon can be found in Ref.~\cite{Bueno:1999ye}.
The liquid Argon TPC is
a very promising detector option which satisfies at best
the above mentioned requisites in terms of granularity, energy resolution
and imaging capabilities if it can be extrapolated to the
relevant mass scale.

In this paper we study the performance of a very massive Liquid Argon TPC as a
nucleon decay detector. 
In the present study, we address for the first time
the effect of the charged cosmic rays background as a function
of the depth (i.e. rock overburden) of the underground detectors.
In particular, we address the possibility to perform proton
decay searches in ``shallow depth'' configurations~\cite{nufact06}.
Since the new large underground detectors will require either a new
site to be excavated or the extension of an existing infrastructure, 
it is important to understand if very-sensitive nucleon decay searches
do necessarily require deep underground locations, like those typically
encountered in existing or planned laboratories~\cite{haxton_nufact06}. 

\begin{figure*}
\centering
\includegraphics[width=0.85\textwidth]{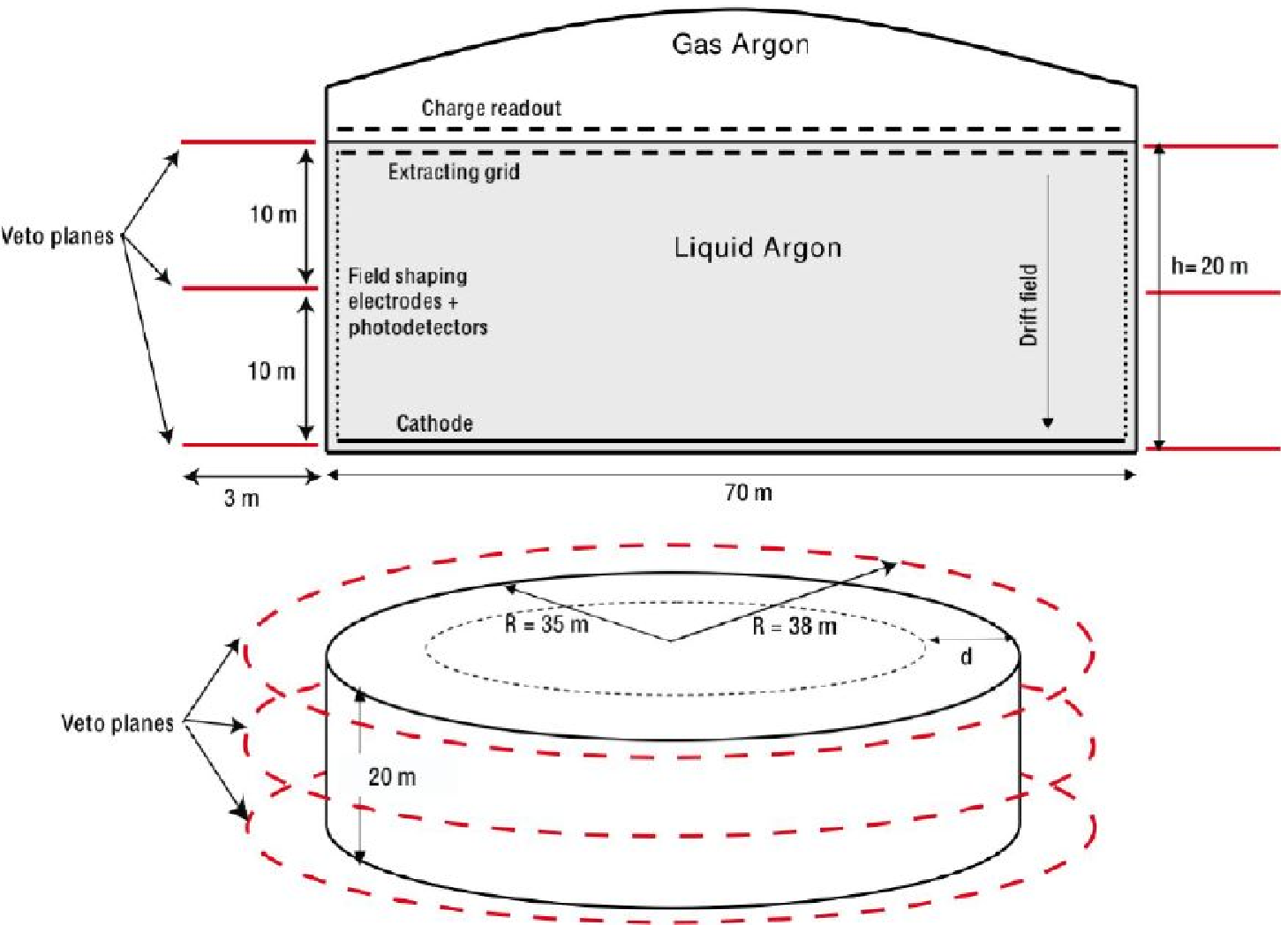}
\caption{Conceptual design of a 100\,kton LAr TPC detector
(see Ref.~\cite{Rubbia:2004tz} for details). To suppress
cosmogenic backgrounds, we consider the 
possibility to locate three large area veto 
planes to detect remaining penetrating cosmic ray muons passing
in the vicinity of the Argon detector (see text).}
\label{fig:s100ktTPC}
\end{figure*}

The paper is organized as follows: 
some considerations about the
 conceptual design of the detector are outlined
 in section~\ref{sec:detcon}.
The discussion of the assumed detector configuration 
and the simulated physics process for signal and background events 
are described in section~\ref{sec:LArTPC}.
The description of the analysis cuts designed to suppress atmospheric neutrino
and muon-induced backgrounds from proton and neutron decay channels
are given in section~\ref{sec:analysis}.
Finally, the results obtained in terms of lifetime sensitivities for the considered
decay modes are given in section~\ref{sec:sensitivity}.

\section{Detector concept}
\label{sec:detcon}
Our analysis assumes the scalable concept of a liquid Argon TPC, 
as proposed in~\cite{Rubbia:2004tz} (see Figure~\ref{fig:s100ktTPC}).
Other designs have been presented in Ref.~\cite{clinesergiamp}. 
An LOI based on a more standard configuration and a surface detector
has also been submitted to FNAL~\cite{Bartoszek:2004si}.

The design of Ref.~\cite{Rubbia:2004tz} relies on 
(a) industrial tankers developed by the petrochemical industry (no R\&D required, readily available, safe) 
and their extrapolation to underground or shallow depth LAr storage,
(b) novel readout methods with embedded charge chain
 for very long drift paths, based on e.g. LEM readout,
(c) new solutions for very high drift voltage,
(d) a modularity at the level of 100 kton
and (e) the possibility to embed the LAr in a magnetic field~\cite{Ereditato:2005yx}. 
Such a scalable, single LAr tanker design is the most attractive solution from the point of view
of physics, detector construction, operation and cryogenics, and finally cost. 

An R\&D program is underway with the aim of optimizing the design~\cite{Ereditato:2005ru}.
Concerning the further consolidation of the technology, we point out that a test-beam dedicated
to the reconstruction and separation of electrons from neutral pions has been discussed~\cite{epilar}.
In addition, a $\simeq 100$~ton liquid Argon TPC to complement the 1~kton Water Cerenkov
detector at the potential 2~km site 2.5$^o$ off-axis from the T2K beam
has also been considered~\cite{t2kprop}. If realized, this unique experimental setup
will allow to compare the performance of the liquid Argon TPC
to the Water Cerenkov ring imaging and to reconstruct neutrino events
directly in the same beam with a statistics of more than 100'000~events per year,
sufficient to extrapolate the atmospheric neutrino background in a potential
1000~kton$\times$year exposure.
ICARUS~T600~\cite{t600paper}, to be
commissioned in the coming years, will
detect too few contained events to demonstrate the ultimate physics performance
of the technology.

When operating TPCs with long drift paths and charge amplification
at the anode as the one considered here, one must pay
attention to possible drift field distortions caused by the space charge
created by the positive ions slowly drifting towards the cathode. Indeed,
it is well known that ion (and hole) mobility in liquid Argon is small
compared to that of electrons ($\mu_{ions}\approx (10^{-2}\div 10^{-3})\mu_e$), 
and inconsistencies among experimental measurements were attributed
to the influence of the motion of the liquid itself~\cite{ionmobs}.

The problem of ions accumulating in the TPC drift volume
has been effectively overcome by introducing an additional grid (`gate')
between the drift volume and the readout plane~\cite{Nemethy:1982uw,Amendolia:1985cq}.
The gate is normally closed and only opens for a short period of time when an
external trigger signals an interesting event. This method is easily applicable
to our design in the case of a pulsed artificial neutrino beam source, as the one considered
in Ref.~\cite{Meregaglia:2006du}. 

In the case of proton decay (or atmospheric
neutrino background) where the detector must be continuously
sensititive, the prompt scintillation light with an appropriate
detection threshold (e.g. $>100$~MeV) could be used to signal
the (rare) occurrence of interesting events. Unfortunately, 
this is impracticable at shallow depths because, as it will be shown later,
 the fiducial volume occupancy is dominated
by remaining cosmic ray events crossing the detector: at
a rock overburden $\lesssim 1$~km w.e., 
the average number of cosmic ray muons crossing the detector within
a drift time is $\gtrsim 1$ (see Table~\ref{tab:nmninLar01}). This situation worsens when
the rock overburden is decreased.

Hence, either the rock-overbuden is $\gtrsim1~$km w.e.
or the TPC must be continuously active and
we must consider space charge effects from accumulating ions
in the liquid Argon volume. 
It is possible to show that asymptotically the stationary density of ions $\rho_i$
in the drift volume of a double phase readout system with charge
amplification is given by:
\begin{equation}
\rho_i(t\rightarrow\infty)\simeq \frac{G \epsilon_{feedback} \eta_{v-l} \epsilon_{grid}
\left<\frac{dE}{dx}\right> x}{\mu_{ions} E W_e(E)} \phi_\mu
\end{equation}
where $G$ is the average gain of the charge readout system,
$\epsilon_{feedback}$ is the efficiency for ions to feed-back into
the gas region below the charge amplification planes,
$\eta_{v-l}$ is the efficiency for ions to be transmitted
from the gaseous phase into
the liquid phase,  $\epsilon_{grid}$ is the
transparency of the extraction grid when traveling from the 
higher extraction field to the lower drift field,
$<dE/dx>$ is the average stopping power of the cosmic ray muons,
$x$ is the average path of the muons in the fiducial volume,
$\phi_\mu$ the flux of cosmic muons at the detector upper surface
per unit surface and per unit time,
$\mu_{ions}$ is the ion mobility, $E$ the electric drift field and $W_e(E)$
the field-dependent average energy expended by a cosmic
ray muon to create an ion-electron pair in the
medium (field dependence arises from columnar ion-electron 
recombination~\cite{Amoruso:2004dy}). 

Assuming a stationary situation,
ionization electrons will drift in a medium homogeneously
filled with the above calculated
density of ions.
The resulting free electron lifetime can be estimated
as $\tau_e=1/\left(k_r(E)\rho_i\right)$
where $k_r$ is the electron-ion recombination rate  (which depends on
the electric field).

We note that $\epsilon_{feedback}\ll 1$,
because of electron diffusion in the amplification gap, as a result of which ions
can follow field lines ending up on the electrodes of the amplification
system rather than on the external field lines; 
this is particularly true for the
devices with multiple stages that we operate~\cite{Kaufmann:2006hp}.
In addition,
$\epsilon_{grid}\lesssim 0.3$ since the extraction field is typ. 3~kV/cm
and the drift field is typ. 1~kV/cm. 

The efficiency $\eta_{v-l}$ for the transfer of positive
ions from the vapour into the liquid Argon phase has, as far as we know, 
not being measured. When the distance between the ion and the liquid-vapor
interface is greater than several Angstroms (the Argon bond length
is $\approx 3.7~\AA$), the liquid may be treated as
a continuum~\cite{bruschi66}. Assuming a planar interface between the
liquid and vapor, the situation is then comparable to a boundary-value
problem with dielectrics and can be solved with the charge-image method
in a single dimension: a point
charge $q$ in the medium  $\epsilon$ at the interface with a medium  $\epsilon'$ 
feels a charge image $q'$ in the medium $\epsilon'$ given by $q'=-(\epsilon'-\epsilon)/(\epsilon'+ \epsilon)q$. For $\epsilon'>\epsilon$, $q.q'<0$ and the force is attractive. 
For $\epsilon'<\epsilon$, $q.q'>0$ and the force is repulsive. 
Given the difference
in dielectric constants between the liquid Argon $\epsilon_l\simeq 1.54$ and
its gas phase $\epsilon_v\simeq 1$, a charge in the liquid will feel a repulsive force
from the interface, while a charge in the gas will be attracted to the interface 
and will tend to ``stick'' at the surface.
The potential energy of the charge $q$ placed at a distance $d>0$ from the planar liquid-vapor
interface 
is~\cite{bruschi66,rayfield71,bruschi75,borghesani90}:
\begin{eqnarray}
V_{l}(d) = \frac{q^2}{16\pi\epsilon_0\epsilon_{l}}
\left(\frac{\epsilon_l-\epsilon_v}{\epsilon_l+\epsilon_v}\right)\frac{1}{d}\equiv \frac{A_l}{d} \\
V_{v}(d) = \frac{q^2}{16\pi\epsilon_0\epsilon_{v}}
\left(\frac{\epsilon_v-\epsilon_l}{\epsilon_l+\epsilon_v}\right)\frac{1}{d}\equiv \frac{A_v}{d}
\end{eqnarray}
where the index corresponds to the charge in the liquid ($l$)
or in the vapor ($v$).  Classically, the potential barrier is infinite 
at the interface ($d=0$) and a quantum mechanical treatment
is needed. In practice, one can also assume that in order to cross the boundary
the charge must pass above a finite, classical potential, whose
height is to be determined experimentally~\cite{rayfield71}.

\begin{figure}
\centering
 \setlength{\unitlength}{1mm}
 \begin{picture}(80,70)
   \put(0,0){\includegraphics[width=82\unitlength]{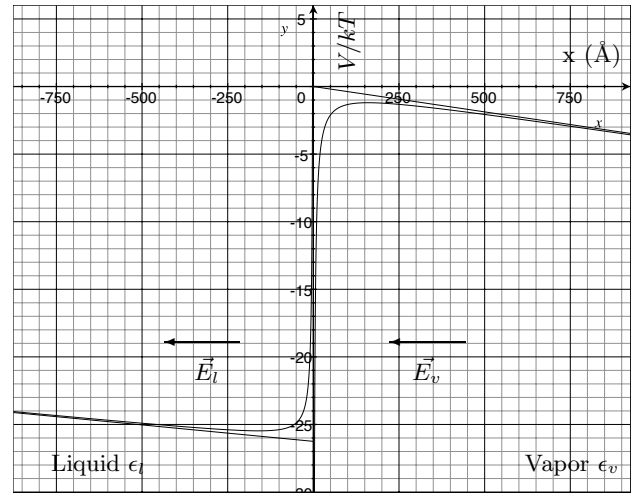}}
   \put(73,60){\makebox(0,0)[tl]{x (\AA)}}
   \put(43,56){\rotatebox{90}{$V/kT$}}
   \put(5,5){\makebox(0,0)[tl]{Liquid $\epsilon_l$}}
   \put(68,5){\makebox(0,0)[tl]{Vapor $\epsilon_v$}}
   \put(30,20){\vector(-1,0){10}}
   \put(24,18){\makebox(0,0)[tl]{$\vec{E_l}$}}
   \put(60,20){\vector(-1,0){10}}
   \put(53,18){\makebox(0,0)[tl]{$\vec{E_v}$}}
 \end{picture}
 \caption{\small Illustration of the potential felt by quasi-free electrons as a function
 of the position $x$ in $\AA$ across the liquid-vapor interface in the presence
 of an external drift field. The potential is given in units of $kT$ 
 where $T=87$~K is the temperature of the liquid Argon at normal
 pressure.}
 \label{fig:potbarriere}
\end{figure}

\begin{figure}
\centering
 \setlength{\unitlength}{1mm}
 \begin{picture}(80,70)
   \put(0,0){\includegraphics[width=82\unitlength]{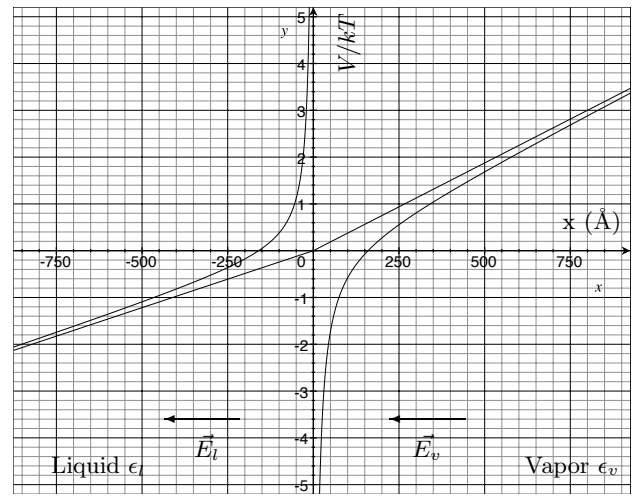}}
   \put(73,38){\makebox(0,0)[tl]{x (\AA)}}
   \put(43,56){\rotatebox{90}{$V/kT$}}
   \put(5,5){\makebox(0,0)[tl]{Liquid $\epsilon_l$}}
   \put(68,5){\makebox(0,0)[tl]{Vapor $\epsilon_v$}}
   \put(30,10){\vector(-1,0){10}}
   \put(24,8){\makebox(0,0)[tl]{$\vec{E_l}$}}
   \put(60,10){\vector(-1,0){10}}
   \put(53,8){\makebox(0,0)[tl]{$\vec{E_v}$}}
 \end{picture}
 \caption{\small Same as Figure~\ref{fig:potbarriere} but for positive ions.}
 \label{fig:potbarrier}
\end{figure}

In the presence of an external difference of potential to create
a drift field, the total potential energy
is $V(x)=V_l(|x|)+qE_lx$ for $x<0$ (liquid phase) and 
$V(x)=V_v(x)+qE_vx$ for $x\ge 0$ (vapor phase), where $E_l$ (resp. $E_v=(\epsilon_l/\epsilon_v)
E_l$) is the induced electric field in the liquid (resp. vapor) phase.

Experimental data has been collected on the extraction
of quasi-free electrons from liquid into gaseous Argon~\cite{borghesani90}.
The potential felt by such electrons is illustrated in Figure~\ref{fig:potbarriere}.
It is plotted in units of $kT$ 
 where $T=87$~K is the temperature of the liquid Argon at normal
 pressure.
 Photoelectric effect measurements indicate
that the electrons in the liquid are shifted by an amount
$V_0 = -0.21$~eV with respect to the energy in vacuum (or vapor)~\cite{tauchert}.
The external drift field and the repulsive potential at the interface
generate a minimum potential in the liquid phase
at a distance $d^l_m=(A_l/eE_l)^{1/2}$ from the interface
with a value $V^l_m=2(A_leE_l)^{1/2}$ and similarly in the
gas phase.
The barrier height as seen from the minimum in the liquid at $x=-d_m$
is therefore 
$\Delta V(E)=V_0-V^l_m-V^v_m = V_0 - 2e^{1/2}\left(A_l^{1/2}E_l^{1/2}+A_v^{1/2}E_v^{1/2}\right)$.
The mean life time to traverse the interface (or trapping time)
can be analyzed with the aid of the Smoluchowski equation~\cite{rayfield71}, 
predicting $\tau_{e}\propto \exp(\Delta V/kT)$. Experimental data however
indicate that the above-calculated thermionic current is increased by a rate
$\propto \mu E/\lambda_1$ where $\lambda_1$ is the
momentum-transfer mean free path, as predicted
by the Shottky model of electric field enhanced thermionic
emission~\cite{borghesani90}, giving $\tau_e\propto\exp\left(\Delta V/kT\right)/E$,
with a stronger influence
of the drift field.

In the case of the ions, the potential for a positive charge 
is illustrated in Figure~\ref{fig:potbarrier}.
The ion is attracted very near the interface and must tunnel through
the repulsive potential existing on the liquid side of the interface in
order to enter the liquid.
A semi-classical approach is not justified. 
The transport of $Tl$ positive ions from xenon vapor to liquid was successfully
reported with a $\simeq 10\%$ efficiency in Ref.~\cite{walters}.
There appears to be a difference in the
behavior of positive ions between the liquid-to-gas and gas-to-liquid
transition~\cite{bruschi66}.
While electrons could traverse the liquid-vapor interface the positive ions
could not. This is consistent with the fact that
the ion mobility in the liquid is too small to allow for the external
field to enhance the thermionic emission.
However, the transmission of negative ions through the liquid-vapor interface of
neon has been observed with trapping time of the order of 10--100~s at fields
of 0.1--1~kV/cm~\cite{bruschi75}. This is explained
by the type of clustering around the different species of particles. If a sort
of bubble surrounds negative charges then the breaking
of the bubble at the surface could let them escape. 
In conclusion, the phenomenon of transfer of the ions through
the interface is expected to be rather complicated,
not well understood and has not been
measured. In the case relevant to this paper, 
it is likely that some fraction of the Argon ions will cross into
the liquid, although possibly with long trapping times,
and experimental studies are needed to assess how much.
For safety, we will conservatively assume full transmission
of Argon ions across the interface.

We note that competing with the process of transfer from the gas to
liquid is the radial drifting of ions at the surface of the liquid-gas
interface towards the edges of the volume, until they reach the field shaping
electrodes. There they are neutralized by the external power supply.

We now compute the free electron lifetime due to the recombination
with the accumulating ions.
In the {\it worst case} considered in this paper corresponding
to the ``under the hill location'' (see Table~\ref{tab:nmninLar01}), 
we contemplate $\simeq1000$~muons
crossing the detector per second. We
conservatively assume
that each of these muons will vertically cross the entire drift region, 
hence $x\simeq h=20$~m,
that the average charge gain is $G\simeq 150$ (see Ref.~\cite{Rubbia:2004tz}), 
very conservatively that
$\epsilon_{feedback}\simeq 0.5$, $\eta_{v-l}\simeq 1$ although we expect this to be a
pessimistic assumption, $\epsilon_{grid}\simeq 0.3$
and that the ion mobility in the range $(0.2 \div1)\times 10^{-3}$~cm$^2$/V/s~\cite{ionmobs}.
These assumptions yield $\rho_i\simeq (0.7 \div 3.7)\times 10^5$~ions/cm$^3$.
 Assuming a recombination rate $k_r = 1\times 10^{-4}$~cm$^3$/s, consistent with 
 measurements at the relevant electric drift fields~\cite{shinsaka},
implies that the electron lifetime due to ion accumulation in the medium is
$\tau_e>30 \div 140$~ms, to be compared with a maximum electron drift
time $\sim 10$~ms.  
We stress that actual values of 
$\epsilon_{feedback}$ and $\eta_{v-l}$ are expected to increase
these values even further.
We conclude that in all configurations considered in this paper,
charge attenuation due to recombination with
ions in the medium is expected to be negligible compared to other charge attenuating 
effects,  like e.g. attachment to electronegative impurities in the Argon~\cite{Amoruso:2004ti}. 

\begin{figure}
\begin{center}
\includegraphics[width=0.45\textwidth]{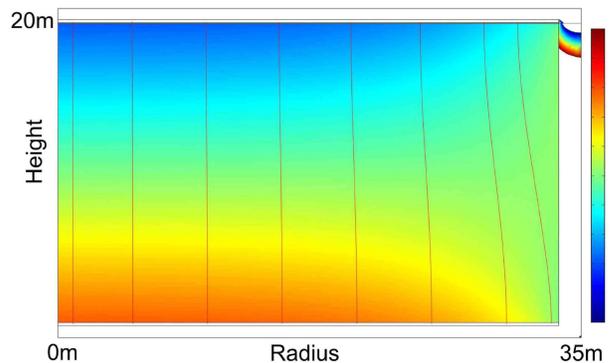}
 \caption{\small Drift field map with an unrealistic ion charge density
 of $2\times 10^5$~ions/cm$^3$=$4\times 10^{-8}$~C/m$^3$ (see text). The field
 distortion is $\pm 30\%$. Dark blue corresponds to 0.5~kV/cm and red
 to 1.5~kV/cm.}
 \label{fig:field1}
\end{center}
\end{figure}

\begin{figure}
\begin{center}
\includegraphics[width=0.45\textwidth]{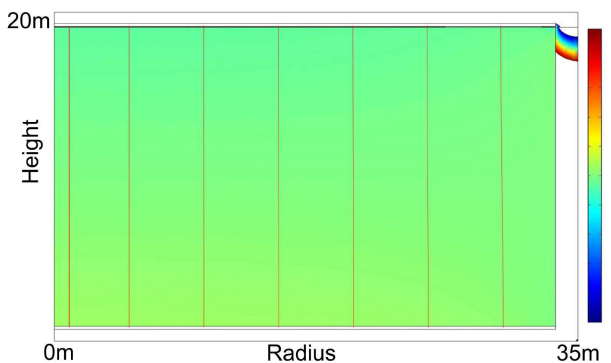}
 \caption{\small Same as Figure~\ref{fig:field1} with a ion charge density
 of $2\times 10^4$~ions/cm$^3$=$4\times 10^{-9}$~C/m$^3$.
 The field distortion is $\pm 3\%$.}
 \label{fig:field2}
\end{center}
\end{figure}

We now turn to the drift field map distortions. The results of numerical simulations
for the electric field
assuming a homogeneous positive charge density distribution in the liquid Argon
fiducial volume of resp.~$\rho_i = 2\times 10^5$~ions/cm$^3$ and
$2\times 10^4$~ions/cm$^3$ and the contributions of
ions directly produced in the ionization process
are shown 
in resp.~Figures~\ref{fig:field1} and \ref{fig:field2}. The cathode is placed at $-2$~MV and
the volume is enclosed by field shaping electrodes.
Dark blue corresponds to 0.5~kV/cm and red to 1.5~kV/cm.
With an unrealistic ion charge density 
of $2\times 10^5$~ions/cm$^3$=$4\times 10^{-8}$~C/m$^3$, the field
is distorted by $\pm 30\%$.  As expected, for a more realistic
charge density of $2\times 10^4$~ions/cm$^3$, field variations
are $\pm 3\%$. The effect of ions directly produced in the
ionization is negligible.

We point out in this context that the realization and successful operation on surface
of a 5 m long detector column~\cite{Ereditato:2005ru}
(ARGONTUBE) will allow to experimentally verify these hypotheses
and to prove the feasibility of detectors
with long drift paths, hence representing a very important milestone in
the conceptual proof of the detector design.

\section{Simulation framework}
\label{sec:LArTPC}

As already pointed out in Section~\ref{sec:intro}, the nucleon decay searches aim at discoveries
at the few event level in an atmospheric neutrino background sample
of more than $10^5$ events. The understanding of this background will
require the collection of a commensurate sample in a neutrino beam of
the relevant energy.

In the meantime, a reliable MC description of all physical effects involved is
important to correctly estimate nucleon decay sensitivities.
Whenever possible, our results are based on
full simulations. Several neutron and proton decay channels
with proper treatment of nuclear effects have been simulated.
Likewise, the two dominant background
sources relevant to our search were simulated in details: 
the atmospheric neutrino background and the cosmic muon-induced background. 
The next sections describe this in more details.

\subsection{Detector geometry and simulation}

We considered as reference detector a homogeneous volume 
of liquid Argon in a single giant volume composed of
a cylindrical volume 
of 20~m height and 70~m diameter for a total
of 100~kton. 
This geometry was implemented in a full simulation based
on the \GEANT{} toolkit~\cite{Allison:2006ve, Agostinelli:2002hh}.

For each event generated within the liquid Argon volume, final
state particles are transported through the medium, with the
possibility of secondary interactions. 
The detector effects have been included in the production
and transport of the events by simulating the liquid Argon response.
For ionizing particles, the
deposited energies in microscopic sub-$mm^3$ volumes of Argon are used to compute the amount of
ionization~\cite{Amoruso:2004dy} and scintillation yields~\cite{Cennini:1999ih}.
Ionization charge is collected in two perpendicularly segmented striped
readout planes with 3~mm pitch. This information is digitized taking into account
the response of a typical potential charge readout electronic preamplifier and is used to create the two
perpendicular charge readout views.
Scintillation light is propagated 
through the medium and collected on photodetectors located on
the inner surface of the detector~\footnote{The scintillation light signal has
not being used in the present analysis, but will in the case of the actual
experiment yield the $T_0$ of the event, allowing to reconstruct the
position of the event along the drift direction.}.

\subsection{Nucleon decay signal simulation}
\label{sec:simula}

Several neutron and 
proton decay channels have been studied (see Section~\ref{sec:analysis}). 
For each channel, 2000 signal
events were generated and fully simulated inside the detector. 
Nucleon decay events are characterized by a definite value of the total
energy and by the fact that the total momentum of the decay products must
be zero. These features, which are true for the decay of a free nucleon, are
only approximately verified for a nucleon bound in a nucleus. 
 
The nuclear effects, the distortions of the energy and momentum distributions
due to Fermi motion (since the recoil
nucleus is not measured), and the reinteraction of decay particles
with the nucleus have been treated with the \FLUKA{}~\cite{Ferrari:2005zk} package.
The treatment is similar to that used for nuclear effects in neutrino interactions,
see e.g. Ref.~\cite{Battistoni:2006da}, and will be used also in the the context
of simulation of atmospheric neutrino background (see Section~\ref{subsec:atmbkg}).

In a given signal event, energy and relative orientation 
of the decay products are selected according to conservation laws and phase
space. The resulting configuration is inserted into the target nucleus and
used as initial step for the \PEANUT~\cite{Ferrari:2005zk} 
simulations. The position of the
decaying nucleon in the nucleus is sampled from a probability distribution
proportional to the density profile for the selected nucleon type. The
Fermi motion of the original nucleon is then sampled according to the local
Fermi distribution. The decay products are followed like any other secondary
particle in \PEANUT, thus exploiting all the details of its nuclear
modeling, as described in the following paragraphs. See Ref.~\cite{Ferrari:2005zk} 
and references therein for full details.

  Nuclear effects in nucleon decays can be roughly divided into those of
the nuclear potential and those due to  reinteractions of decay products. 
Bound nucleons in nuclei are subject to a nuclear potential. The
Fermi energy (or momentum) must be calculated from the bottom of this nuclear
potential well, and the removal of a nucleon from any stable nucleus 
is  always an endothermic reaction. 
When a nucleon decays, some energy is spent to take it
out of this well: the minimum energy is given by the nucleon separation
energy (around 8 MeV), and corresponds to the decay of a nucleon at the
Fermi surface. In this case, the daughter
nucleus is left on its ground state.  More deeply bound nucleons 
decay leaving a hole in the Fermi sea, that corresponds to an excitation
energy of the daughter nucleus, and an additional loss of energy of the
decay products. This energy is then spent in evaporation and/or gamma
deexcitation. Thus, the invariant mass of bound nucleon decay products is
expected to be always slightly smaller than the free nucleon mass,
and spread over a range
of about 40 MeV.  Correspondingly, the Fermi momentum of the decaying
nucleon is transferred to the decay products and compensated by the recoil
of the daughter nucleus. Additional momentum distortions come from the 
curvature of particle trajectories in the nuclear potential.
 
  Reinteractions in the nuclear medium also play an important role. Decay
products can lose part of their energy in collisions, or even be absorbed 
in the same nucleus where they have been created. This is particularly true
for pions, that have an important absorption cross section on nucleon
pairs, while kaons have smaller interaction probability.    

Nucleon-nucleon total cross sections, both elastic and inelastic, used
in \FLUKA{} are
taken from available experimental data. Elastic scattering is
explicitly performed according to the experimental differential cross
sections. Pion induced reactions are more complex, mainly
because of two- and three-nucleon absorption processes.
Above the pion production threshold, the inelastic interactions are
handled by the resonance model.
Other pion-nucleon interactions proceed through the non-resonant
channel and the p-wave channel with the formation of a $\Delta$
resonance, with modified resonance parameters taking into account
nuclear effects.

The conservation of strangeness leads to very different interactions of the 
K$^+$, K$^0$ and K$^-$, 
$\bar{\mathrm K}^0$ with nucleons at low energies. K$^-$
have a large cross section for hyperon production, with the $\Sigma \pi$
and $\Lambda \pi$ channels always open. 
A detailed treatment of $K^+$ interactions has been 
developed in \PEANUT, whereby the $K^+$ nucleon system is described by 
phase shift analysis. This treatment was used in the present
analysis, where it was very relevant for the nucleon decays
involving kaons. In particular, $K^+$ involved in proton decays are below 700~MeV/c,
where the inelastic contributions are very small and elastic scattering 
does not affect strongly the efficiency. Charge exchange $K^++n\rightarrow K^0+p$
can lead to a loss of efficiency.
In the simulations of $p\rightarrow \bar\nu K^+$ decays, 
less than 4\% of the kaons were lost due to nuclear processes (see Section~\ref{sec:analysis}).
In comparison, in the simulated $p\rightarrow e^+\pi^0$ decays, the neutral pion
is absorbed within the nucleus with a probability of $\simeq 50$~\%.

In the treatment of nuclear reinteractions, the cross-sections are 
modified to avoid too short mean free paths in nuclear matter
by taking into account Pauli blocking, 
antisymmetrization of fermions, hard-core effects and
 formation zone or coherence length for non-fragmenting processes.
The intra-nuclear cascade step goes on until all nucleons are below 50~MeV 
and is followed by a pre-equilibrium emission phase in which additional
nucleons can be emitted.

Many benchmarks showing good agreement between \PEANUT\ results and
nuclear reaction data can be found in the main
Ref.~\cite{Ferrari:2005zk}.

\subsection{Atmospheric neutrino background}
\label{subsec:atmbkg}

The atmospheric neutrino flux has been computed by several groups.
We take results from the \FLUKA{} group~\cite{Battistoni:2002ew} and 
the HKKM 2004~\cite{Honda:2004yz} (Honda) models. 
The \FLUKA{} model is estimated to have a 7\% uncertainty for the
primary spectrum, 15\% for the interaction model, 1\% for the
atmosphere profile, 2\% for the geomagnetic field and a total 
17\% uncertainty.
The HKKM model has about 10\% uncertainty when the neutrino energy is below 10~GeV,
but the uncertainties are still large when the neutrino energy is above 10 GeV,
due to the uncertainties of the primary cosmic ray flux and the interaction model above 100 GeV.

The neutrino-cross
sections and the generation of neutrino interactions are based on the {\tt NUX}~\cite{NUX}
code. This code can be
used in fixed target mode for incoming 
neutrino energies from $\simeq 20$~MeV up to 10~TeV.
{\tt NUX} can generate neutral current (NC) and charged current (CC) processes, and
includes all basic processes to properly describe neutrino
interactions in different kinematical regimes. In particular, a distinction is made
among the following processes: quasi-elastic process (QE), inelastic process (DIS)~\footnote{The
DIS process includes single-pion production via baryon resonance excitation within a 
parton model formalism, inspired from the concept of duality.} 
and charm production.
These processes are appropriately matched to reproduce the total inclusive and
existing exclusive experimental data to the best of current experimental knowledge.
The mass of the outgoing lepton and of the target
nucleon are taken into account in the kinematics and in the differential cross-section.
In order to take into account nuclear effects, an interface with \FLUKA{} was
implemented, yielding the so-called {\tt NUX-FLUKA} generator: a primary nucleon is chosen 
according  to the density profile for the selected nucleon type and the
Fermi motion of the original nucleon is sampled according to the local
Fermi distribution, just as in the case of the nucleon decay simulation
(see Section~\ref{sec:simula}).
A neutrino interaction is generated and the
final state intra-nuclear cascade is taken into account by propagating
final state particles through the nucleus with
\PEANUT. 

The {\tt NUX-FLUKA} model was
benchmarked with the data of the NOMAD~\cite{NOMAD} experiment
at high neutrino energies ($>5$~GeV), 
and was compared at low neutrino energies ($<5$~GeV) with other available generators
like  {\tt NUANCE}~\cite{Casper:2002sd} and was shown to give very 
similar particle multiplicities and distributions~\cite{compnuance}.
The production of strange particle at high energy was studied in NOMAD~\cite{Astier:2001vi}
and found to be properly described after tuning of the {\tt JETSET}~\cite{Sjostrand:1986hx} fragmentation parameters. At low energy, the production of kaons primarily
proceeds through the decay chain of baryon resonances. In this regime,
the yield of kaon is very strongly suppressed compared to that of pions.

\begin{table}
    \begin{center}
\begin{tabular}{|c|ccc|ccc|ccc|ccc|}
\hline
Atmospheric & \multicolumn{3}{|c|}{$\nu_e$ CC (evts/kt/year)} & \multicolumn{3}{|c|}{$\bar\nu_e$ CC (evts/kt/year)} \\
 Flux Model  & DIS     &     QE     &    Tot   &   DIS   &   QE    &    Tot    \\
\hline
FLUKA& 22.9 &   49.0  &  71.9 & 7.2  & 8.4  &   15.6 \\
 ($\bar{E}$(GeV))          & (2.83)& (0.62)&          & (3.12)& (0.79) &          \\
HKKM& 22.7 &   45.6 &  68.3 & 7.2 &  7.8 &   15.0 \\
($\bar{E}$(GeV))          &(3.01) &  (0.64) &            &(3.42) &(0.85)&         \\
\hline
\hline
Atmospheric & \multicolumn{3}{|c|}{$\nu_\mu$ CC (evts/kt/year)} & \multicolumn{3}{|c|}{$\bar\nu_\mu$ CC (evts/kt/year)}\\
 Flux Model  & DIS     &     QE     &    Tot   &   DIS   &   QE    &    Tot  \\
\hline
FLUKA & 42.0 & 80.1 & 122.1 & 15.3 &  18.5 &  33.8\\
 ($\bar{E}$(GeV))          & (4.53)& (0.72)&          & (4.89)& (0.90) &          \\
HKKM & 43.3 & 74.0 & 117.3 & 15.8 &  17.4 &  33.2\\
 ($\bar{E}$(GeV))          & (4.78)& (0.77)&          & (5.19)& (0.98) &          \\
\hline
\hline
      Atmospheric& \multicolumn{3}{c|}{$\nu$ NC (inelastic)} & \multicolumn{3}{c|}{$\bar\nu$ NC (inelastic)} \\
      Flux  Model& \multicolumn{3}{c|}{(evts/kt/year)}      &   \multicolumn{3}{c|}{(evts/kt/year)}    \\
      \hline     
      FLUKA   &     \multicolumn{3}{c|}{23.2}         &   \multicolumn{3}{c|}{9.0}  \\
      ($\bar{E}$(GeV)) &    \multicolumn{3}{c|}{ (3.73)}        &   \multicolumn{3}{c|}{(4.08)} \\
      HKKM   &     \multicolumn{3}{c|}{23.4}         &   \multicolumn{3}{c|}{9.1}  \\
      ($\bar{E}$(GeV))&     \multicolumn{3}{c|}{(3.97)}        &   \multicolumn{3}{c|}{(4.40)} \\
        \hline
    \end{tabular}
\caption{QE, DIS event rates per kton per year, and average energies for FLUKA 2002 and HKKM 2004 flux models and {\tt NUX} neutrino cross-sections. In order to reach nucleon decay sensitivities in
the range of $10^{35}$~years, exposure in the range of 1000~kton$\times$ year will be considered.}
    \label{tab:ratenc}
  \end{center}  
\end{table}

 \begin{figure*}
\begin{center}
\includegraphics[width=0.62\textwidth]{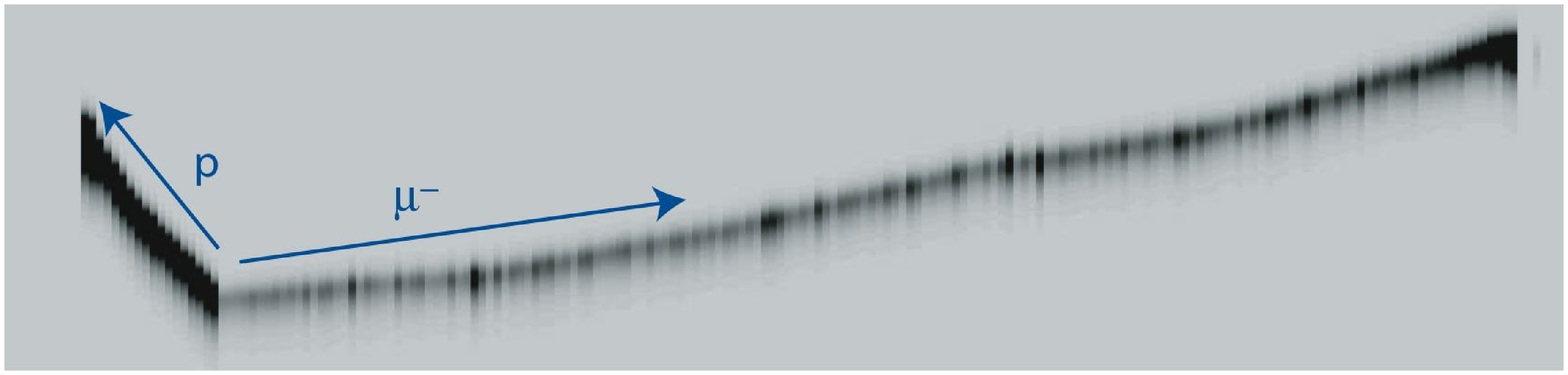}
\includegraphics[width=0.62\textwidth]{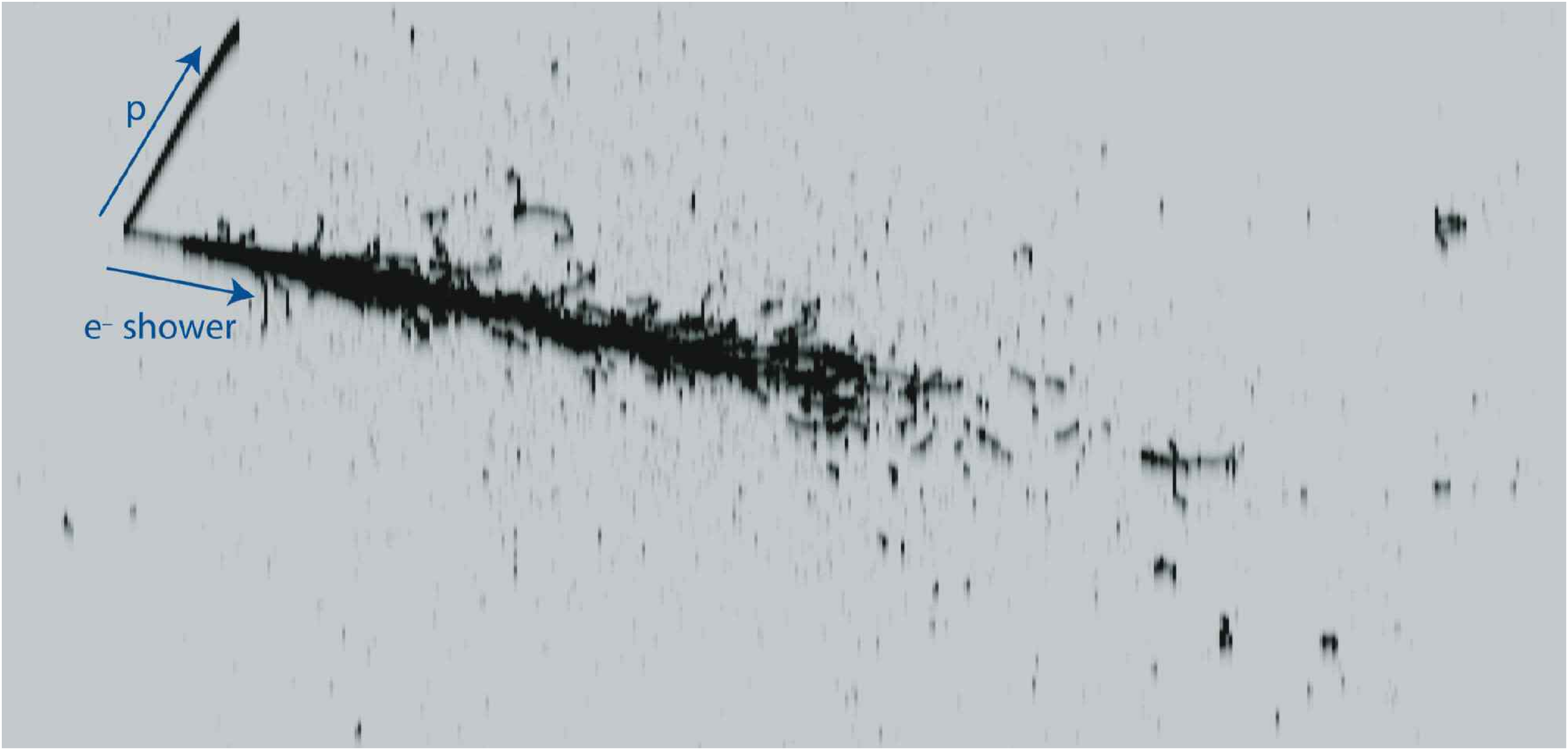}
 \caption{\small Typical $\nu_{\mu}$ and $\nu_{e}$
 QE event in liquid Argon detector ($\nu_{\mu} + n \to p + \mu^-$ and $\nu_e + n \to p + e^-$).}
 \label{fig:TypNuMuQE}
\end{center}
\end{figure*}

\begin{figure*}
\begin{center}
\includegraphics[width=0.5\textwidth]{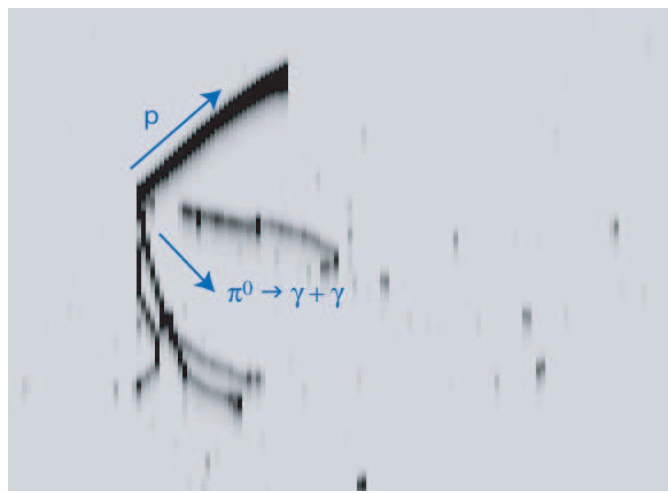}
 \caption{\small Typical $\nu_{\mu}$ NC event in liquid Argon detector ($\nu_{\mu} + p \to \nu_{\mu} + p + \pi^0$).}
 \label{fig:TypNuMuNC}
\end{center}
\end{figure*}

Table~\ref{tab:ratenc} reports the computed number 
of atmospheric neutrino events
per neutrino flavor and per process type, normalized to an
exposure of 1~kton$\times$year.
In order to reach nucleon decay sensitivities in
the range of $10^{35}$~years, exposures in the range of 1000~kton$\times$year will be considered.
Therefore, the simulated data sets correspond to a total exposure of
1000~kton$\times$year or about 250'000~background events. 
Part of this atmospheric neutrino background sample was fully simulated and fully
reconstructed.
Typical atmospheric neutrino events in Argon viewed by
two-dimensional charge readout planes with a pitch of 3~mm
are shown in Figures~\ref{fig:TypNuMuQE} and \ref{fig:TypNuMuNC}. 
Full reconstruction of atmospheric neutrino events
at low multiplicity is rather straight-forward.
In particular, quasi-elastic events reconstruction is discussed in Ref.~\cite{thesisge}.
Actual performance on real quasi-elastic events data can be found in Ref.~\cite{Arneodo:2006ug}.
See also Ref.~\cite{Battistoni:2006nz} for a discussion of precision atmospheric neutrino measurements
with a liquid Argon TPC.

In order to have the largest sample of events to estimate the backgrounds after
selection cuts, a fast reconstruction program was developed which used
Monte-Carlo tracks as input for pattern recognition. Energy and angular resolution
of the detector were properly taken into account by means of the full
simulation. On the other hand, particle reconstruction efficiency was introduced as momentum
thresholds which depended on the particle type. These latter were 
approximately 5~MeV/c for electrons, 10~MeV/c for photons, 20~MeV/c for
muons and pions, 30~MeV/c for kaons, and 300~MeV/c for protons.
Checks were performed to ensure
that the fast simulation reproduced the general features of the background
events on the subsample of fully reconstructed events.

\subsection{Cosmic muon-induced background}
\label{subsec:muonbkg}

A second relevant source of background comes from particles produced by
cosmic muons crossing and interacting in the liquid Argon volume, or 
muons avoiding the active liquid Argon but interacting in the vicinity of the detector. 
The main physical parameter determining the rate of this background
is the rock overburden. 

The cosmic muon intensity as a function of underground depth is obtained 
from Crouch's~\cite{Crouch} fit to the world data and further updated by
the Particle Data Group~\cite{Yao:2006px}. This parameterization agrees
to better than 6\% with the most recent MACRO measurements~\cite{Ambrosio:1995cx}
at Gran Sasso. However, it cannot be applied to the case of a  
shallow depth detector. For this purpose, we generated cosmic $\mu^+$'s and $\mu^-$'s 
according to double differential distributions found in
Ref.~\cite{Yao:2006px} on a large hemisphere at the surface of the Earth
and transported them through rock. In the simulation, all physical processes
were taken in account, including multiple scattering and energy losses.

We aimed at giving here the general trend and the magnitude of the
background, while the precise determination of the muon induced background 
will  depend on many site-specific features: (a) rock chemical composition,
(b) exact distribution of rock overburden (c) 
detector configuration, cryostat, distance to rock, ...

Many relevant processes depend on rock chemical composition.
For example, the neutron production per unit of rock mass
depends on the average atomic number. 
The expected dependence according to Ref.~\cite{Boulbyflux}
is $dN \propto A^{0.76}$, i.e., that means a difference of about 25\% in neutron production 
going from standard rock ($\langle A \rangle = 22$) to the salt rock  ($\langle A \rangle = 30$). 
Particle  propagation through rock of secondary particles produced
in muon interactions is also influenced 
by chemical composition, in first approximation with the 
same behaviour as the nuclear interaction length.

The angular distribution of muons underground depends on the 
actual distribution of rock overburden. If the Earth surface above the site 
is approximately flat, the muon intensity at an angle $\theta$ 
is given by the vertical intensity 
calculated at  a depth $h^\prime=h/\cos{\theta}$ multiplied by a factor $1/\cos\theta$
from the production probability in the atmosphere~\cite{Crouch}.  
If, however, the  detector is located in a mountain or under a hill, the muon intensity 
is determined by the actual rock overburden along any given direction. 
The most important effect is a different total muon intensity with respect to 
the flat surface case, 
and a secondary effect is the difference in veto efficiency (see Section~\ref{sec:RPC})
due to the different muon directionality. Indeed, the deeper the rock overburden,
the more vertical the surviving muons.

Since, as we will show in the following paragraphs, secondary particles that enter
the detector are produced in the first one or two meters of material around it, 
the actual material distribution/composition 
of the LAr container may affect the background and the veto conditions.
This effect has been neglected at this stage and we have assumed rock composition
around the fiducial liquid Argon volume.

\begin{figure*}
\begin{center}
\includegraphics[width=0.95\textwidth]{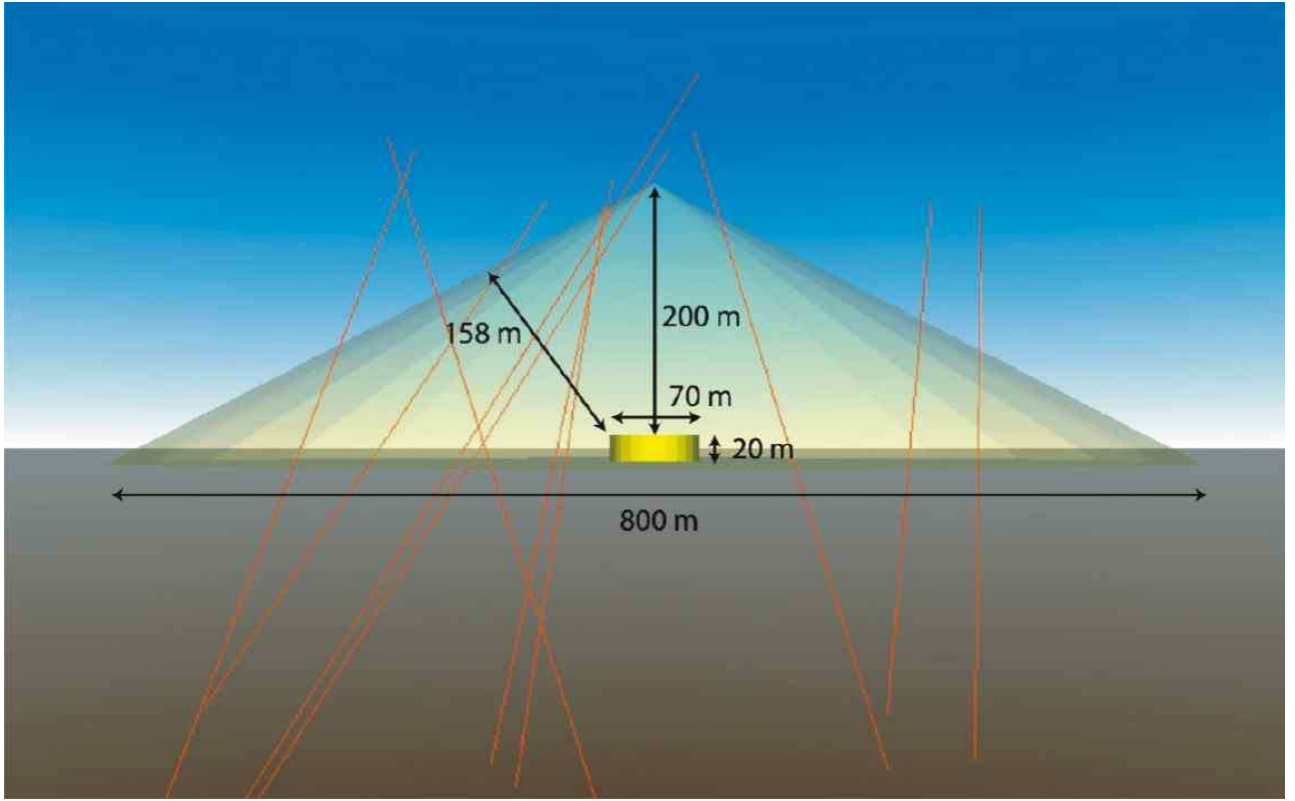}
\caption{\small Under the hill detector configuration. The rock overburden at the vertical
of the detector is 200~m and the shortest overburden is 158~m. This geometry has been
implemented in a full simulation based on \GEANT.}
 \label{fig:geometry3d}
\end{center}
\end{figure*}

We have considered different underground depths and
two different geometries (see Tables~\ref{tab:nmninLar01} and \ref{tab:neutcomparison}) in order to
explore possibilities at shallow depths: (a) under flat ground cover
(e.g. a green-field site~\footnote{If the rock is well known and of very good quality in a given site,
a new hole in virgin ground can be considered. See~\cite{nufact06} and references therein.} 
or a mine with vertical access) at resp. approximately
3, 2, 1, 0.5, 0.13~km water equivalent,
corresponding to resp. 1130~m, 755~m, 377~m, 188~m and 50~m of standard rock;
and (b) detector under a hill configuration at shallow depth (see Figure~\ref{fig:geometry3d}).
A green-field site, or a shallow-depth site under the hill were considered 
in the context of a synergy with an upgraded CNGS accelerator neutrino physics
program~\cite{Meregaglia:2006du}.

\begin{table*}
\centering
\begin{tabular}{|l|c|c|c|c|c|c|c|}\hline
\multicolumn{2}{|c|}{Depth} & Code  &\multicolumn{2}{|c|}{All muons} & \multicolumn{2}{|c|}{$E_\mu>1\ \mathrm{GeV}$} 
& Effective mass \\
Water equiv. & Standard rock & & Particles/s & Particles/10\,ms & Particles/s & Particles/10\,ms &  \\ \hline
\multicolumn{2}{|l|}{Surface detector} & \FLUKA & $1700000$ & $17000$& 1300000& 13000 & -- \\ \hline
$\simeq 0.13$ km w.e. & $50\ \mathrm{m}$ & \FLUKA&  11000 &  110& 10000& 100 & 50~kton\\ \hline
$\simeq0.5$ km w.e.& $188\ \mathrm{m}$ & \FLUKA&  $330$ & 3.3 & $320$& 3.2 & 98~kton\\ \hline
&   $200\ \mathrm{m}$& \GEANT&  -- & -- & $420$& 4.2& 98~kton\\ \hline
$\simeq1$ km w.e.& $377\ \mathrm{m}$ & \FLUKA&  $66$ &0.66 & $65$& 0.65 & 100~kton\\ \hline
$\simeq2$ km w.e.& $755\ \mathrm{m}$ & \FLUKA&  $6.2$ & $0.062$ & 6.2& 0.062& 100~kton \\ \hline
$\simeq3$ km w.e.& $1.13\ \mathrm{km}$ & \FLUKA&  $0.96$ & 0.01& $0.96$&0.01& 100~kton\\ \hline
\hline
\multicolumn{2}{|l|}{Under the hill (see Figure~\ref{fig:geometry3d})} & \GEANT& -- & -- & 960 & $9.6$ & 96~kton\\ 
\hline
\end{tabular}
\caption{Computed average number of muons entering the detector per unit time
for various geographical configurations. The effective mass corresponds to the mass
of Argon that can be used when, in both 2D readout views,
 a slice of size 10~cm around each crossing muon is vetoed.}
\label{tab:nmninLar01}
\end{table*}

The simulations of muon-induced background were performed using \FLUKA{} and 
\GEANT{} toolkit  and the results were compared.
They were carried out in three steps:
(1) in the first step we estimated the number of remaining cosmic muons
crossing the detector or its vicinity per given unit time
as a function of the detector's depth or its geographical configuration; 
(2) then we obtained particle fluxes at the detector surface
from interactions of  those cosmic ray muons in the surrounding rock as a function
of the detector's depth;
(3) the resulting
neutron, kaon and lambda differential flux spectra, rescaled to the rate
expected at a given depth, have then been used to simulate particle
interactions inside the detector. 

As far as \FLUKA{} simulations are concerned, the cross section of the photo-nuclear interaction of high-energy muons is taken from Ref.~\cite{Bezrukov} and validation of the \FLUKA{}
models for muon transport and muon photo-nuclear interactions can be found
in Refs.~\cite{Boulbyflux,FlukaWang}.

In the case of the detector under the hill,
the topology of an existing location was studied~\footnote{The geography of the chosen location
was estimated with the Google Earth tool. See {\tt http://earth.google.com}.} and a simple but conservative description
of the rock profile was implemented in a \GEANT{} simulation. The parameters of the
geometry in a 3D image from the \GEANT{} simulation are shown in Figure~\ref{fig:geometry3d}.
The assumed rock overburden at the vertical
of the detector is 200~m and the shortest overburden is 158~m. 
Cosmic ray muons were simulated on a circular area with a radius of 400~m at a height of 200~m above the detector which corresponds to the top of the hill.

The first concern is the average number of muons entering the detector within a time interval
of 10~ms, which corresponds
 to the assumed time for a full drift over 20~m at a field of 1kV/cm~\cite{Rubbia:2004tz}.
 The results are summarized in Table~\ref{tab:nmninLar01}.
 
 The results from  \FLUKA{} and \GEANT{} were compared considering a flat Earth surface 
 profile and a detector at $\simeq 200$~m underground: muon rates agreed within a factor of $\sim 1.3$.
At depths deeper than about 1~km~w.e. the rate of crossing muons is less than 1 per
10~ms.
Shallower depths e.g. at rock overburdens less than 200~m 
are disfavored because of detector occupancy (at 50~m rock overburden
the average rate of muons with more than 1~GeV crossing in 10~ms is 100).
If we assumed that a slice of size 10~cm in a 2D readout view of the detector
around each crossing muon cannot be used for physics, then the rate at resp.
50~m, 188~m would correspond to an available Argon mass of resp.
50~kton, 98~kton. We do not at this stage consider surface
operation, as proposed in Ref.~\cite{Bartoszek:2004si}.
For the under the hill configuration, more than a million muons, corresponding 
to 156 drifts of 10~ms, were simulated and it was found that on average 9.6~muons enter the Argon
volume in 10~ms. 
This is a factor $\sim 2$ higher than at an equivalent depth
under flat surface, which is still very tolerable.
 
\begin{table*}
\begin{center}
\begin{tabular}{|c|c|c||c|c|c|c|c|}
\hline
Configuration/  &  & Average number $\mu$'s& \multicolumn{2}{|c|}{Neutrons}& \multicolumn{2}{|c|}{Neutral kaons} &Lambdas\\
Depth& Simulation& entering LAr &per year & per $\mu$ in LAr  &per year & per $\mu$ in LAr & \\
& & per 10~ms &   & per 10~ms & & per 10~ms&\\
\hline
$\simeq 0.5$~km w.e.  & \FLUKA{} & 3.3 & 1.9$\times 10^6$ & 1.8$\times 10^{-4}$ & 4500 & 4.3$\times 10^{-7}$ &  $\approx 0.04\times N_{K^0}$ \\
(188~m rock) & & & & & & &  \\
\hline
$\simeq 1$~km w.e.  & \FLUKA{} & 0.66 &  5.5$\times 10^5$ & 2.6$\times 10^{-4}$ & 1300 & 6.2$\times 10^{-7}$ &  $\approx 0.05\times N_{K^0}$\\
(377~m rock)& & & & & & & \\
\hline
$\simeq 3$~km w.e.  & \FLUKA{} & 0.01 & 1.1$\times 10^4$ & 3.6$\times 10^{-4}$ & 25 & 8.2$\times 10^{-7}$ & $\approx 0.06\times N_{K^0}$ \\
(1.13~km rock)& & & & & & & \\
\hline
\hline
Under the hill & \GEANT{} & 9.6& 9.7$\times 10^6$ & 3.2$\times 10^{-4}$ & 1.2$\times 10^3$& 4.0$\times 10^{-8}$ &  -- \\
(see Figure~\ref{fig:geometry3d})  & \FLUKA{} rescaled& & &  & $\approx$1.2$\times 10^4$& $\approx$4.0$\times 10^{-7}$ &  $\approx 0.05\times N_{K^0}$ \\
\hline
\end{tabular}
\vspace{0.1cm}
\caption{\small Cosmogenic background as a function of assumed
depth:  estimated number of neutrons (with kinetic energy above 20~MeV),
neutral kaons and lambda's entering the detector per year, and normalised on the average number of muons entering the active volume per 10~ms, produced in cosmic muon
interactions (hadrons accompanied by a detected muon 
inside LAr imaging have been vetoed).
}
\label{tab:neutcomparison}
\end{center}
\end{table*}

We now focus on the second step where the rate and energy spectrum of 
remaining cosmic ray muons at the underground locations is used to study the 
number of particles entering the detector.
We assume that events in which the parent muon enters 
(before or after the photo-nuclear interaction)
the active LAr volume can be discarded thanks a veto based on the liquid Argon imaging.
This leads us to restrict the background sources to neutral hadrons and,
in particular, to {\it neutrons}, {\it neutral kaons} and {\it lambdas}, produced either
directly in muon photo-nuclear interactions or as secondary products
in hadronic showers in the materials (e.g. rock)
surrounding the detector. We have so far conservatively neglected correlations between
neutral hadrons penetrating inside the LAr in conjunction with (visible)
charged hadrons.

\begin{figure}
\centering
\includegraphics[width=0.49\textwidth]{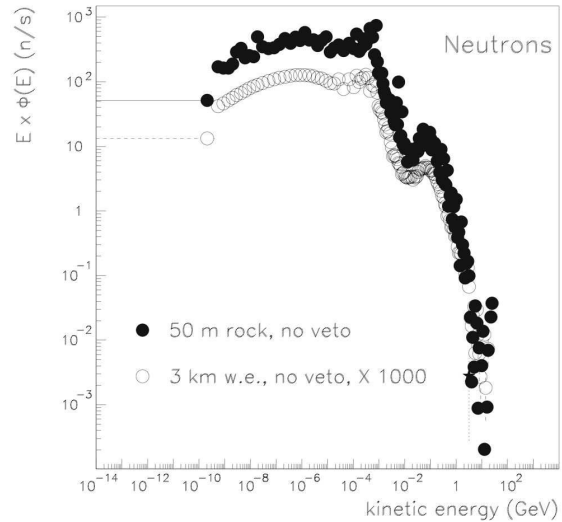}
\caption{Energy distribution of neutrons entering the LAr volume
as predicted by \FLUKA.
\label{fig:nspec} }
\end{figure}

\begin{figure}
\centering
\includegraphics[width=0.49\textwidth]{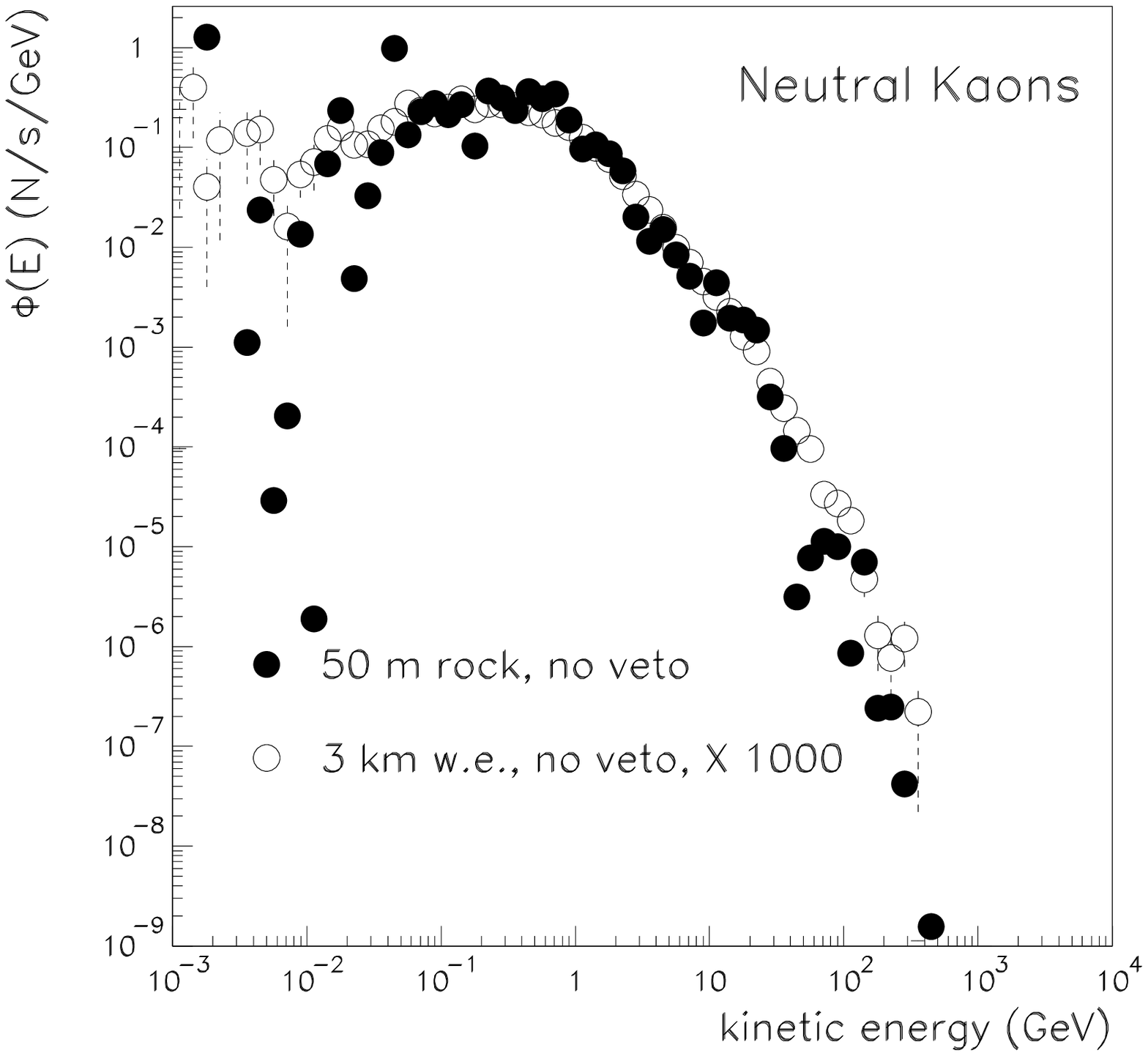}
\caption{Energy distribution of neutral kaons entering the LAr volume
as predicted by \FLUKA.
\label{fig:kspec}}
\end{figure}

\begin{figure}
\centering
\includegraphics[width=0.49\textwidth]{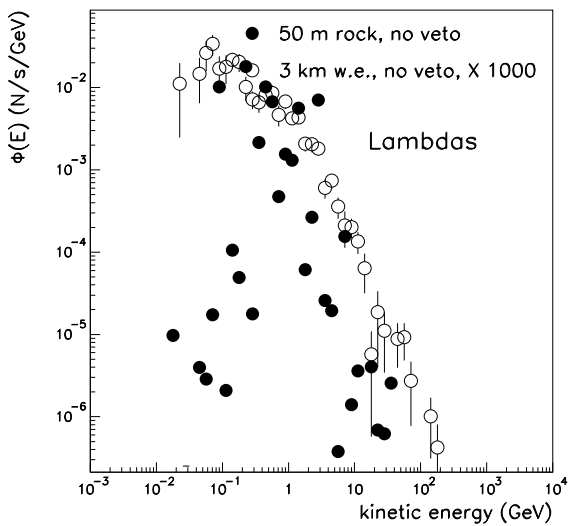}
\caption{Energy distribution of lambda particles 
 entering the LAr volume
as predicted by \FLUKA.
\label{fig:lspec}}
\end{figure}

Neutral hadronic particles have been scored when entering the detector.
The energy-integrated number of particles entering the detector per year
of exposure at some depths are reported in Table~\ref{tab:neutcomparison}.
Neutron, neutral kaon and lambda energy spectra on the detector surface are plotted in
Figures~\ref{fig:nspec}, \ref{fig:kspec} and \ref{fig:lspec}. Their shapes are practically
independent of the depth of rock overburden.
Since low energy neutrons are not a background source for nucleon decay
searches, we will consider only neutrons with kinetic energy above
20~MeV. 


As expected, the particle yields per muons entering the detector, 
increase with depth, since the average muon
energy increases as well. 
The neutrons yields in the \FLUKA{} and \GEANT{} simulations are consistent within a factor of 2. 
As far as neutral kaons are concerned, there is a difference of a factor of 10.
In the case of lambda's, they were not found in the \GEANT{} simulation while in \FLUKA{} they
are expected at the level of 0.04--0.05 times the rate of neutral kaons.
In the following, we will use the yields from \FLUKA{} to estimate backgrounds.

In the third phase, neutrons, neutral kaons and
lambda's were generated with their expected energy
spectrum and were transported in LAr until they underwent an inelastic interaction or decayed. 
These fully simulated interactions were subjected to the selection
cuts and were used to estimate cosmogenic backgrounds (see Section~\ref{subsec:cosmicbkg}).

\subsection{An active very large area cosmic muons veto}
\label{sec:RPC}
\begin{figure}
\begin{center}
 \epsfysize=8.0cm\epsfxsize=8.0cm \epsffile{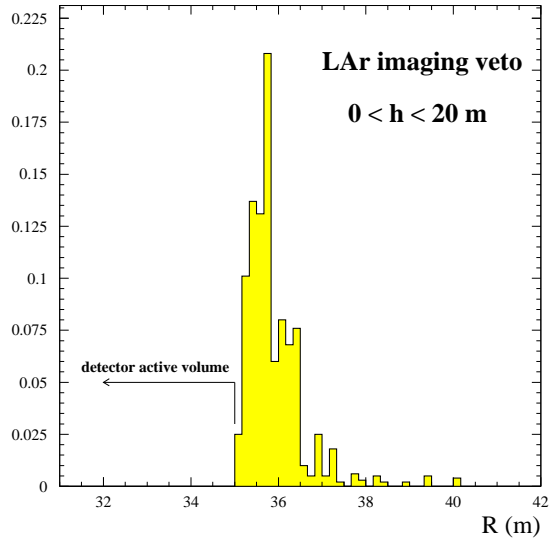}
 \caption{\small
  Radial distribution of photo-nuclear interaction vertexes producing a neutron that
  enters the detector at a depth of 3~km w.e.}
 \label{fig:rzeta}
\end{center}
\end{figure}

The spatial distribution of photo-nuclear interactions in a 3~km~w.e. depth from which 
at least one neutron enters the detector is plotted in Figure~\ref{fig:rzeta}.
Most interactions occur along the detector vertical wall, at an average distance
of one meter from it. The possibility to veto cosmic muon passing in the
vicinity of the detector has therefore been investigated. 

We considered annular planes of active muon detectors with an inner radius of 35~m and an outer radius of 38~m, to detect muons that pass within 3~m from the LAr active volume (also
here we have neglected the details of the liquid Argon tank). Two configurations have been taken into account: 
(a) two planes at the top and at the bottom of the LAr detector; (b) three planes as shown in 
Figure~\ref{fig:s100ktTPC}. 
In the simulations no assumptions are made regarding
their composition, they have a thickness of 1 cm and 100\%
efficiency for  muon crossing the counters. In practice, they could be realized
with Resistive Plate Chambers (RPC)~\cite{Santonico:1981sc,Bellazzini:1985tk} or Glass RPCs (GSC)~\cite{Gustavino:1992fc}. 
The total area needed is about 3800~m$^2$ per plane. 
In the MONOLITH proposal~\cite{monolith} the total active area was envisaged
to be $\approx 54000$~m$^2$. Therefore GSC had to be designed in order to allow large-scale production. 
For techniques on the industrial large area production of glass RPC's
see e.g. Ref.~\cite{Trinchero:2003be}.  

From the simulation 
in the under the hill configuration, we obtained that about 60$\%$ of the muons enter the veto region (i.e. a volume 20~m high and 3~m wide outside the LAr detector) from the top. This value was used as a normalisation to calculate the fraction of muons detected with the two veto configurations, and the results are stated in Table~\ref{tab:mucontained}. Even with the two veto planes configuration, more than 80$\%$ of the muons could be detected.
In conclusion, an active very large area veto composed of 2 or 3 planes could suppress cosmogenic
background by an order of magnitude. In the flat Earth surface geometry or at larger depths, we expect the suppression to be
even stronger since muons are more vertically distributed than in the hill configuration.
These results will be used in the next section to compute sensitivities to nucleon decays.

{\small 
\begin{table}
\begin{center}
\begin{tabular}{|c|c|c|c|}
\hline
& \multicolumn{3}{c|}{Muons (1.6$\times10^6\mu$ simulated per configuration)} \\
  Configuration   & in 1$^{st}$ veto plane & in veto volume & $\%$ detected \\
\hline
Full volume &  228   &377 & 100$\%$\\
\hline
3 veto planes  & 231 &  361& 94$\%$\\
\hline
2 veto planes  & 243  & 343 & 85$\%$\\
\hline
\end{tabular}
\vspace{0.1cm}
\caption{\small Fraction of muons entering the veto volume and detected using veto planes in different configurations. The veto volume is defined as a cylinder 20~m high and 3~meter wider in radius than
the LAr volume.}
\label{tab:mucontained}
\end{center}
\end{table}
}

\section{Nucleon Decay Analysis}
\label{sec:analysis}

The list of studied decay channels in summarized in Table~\ref{tab:channels}.

\begin{table*}
\centering
 \begin{tabular}{|l|c|c|c|c|c|}
\hline
\multicolumn{1}{|c|}{} & \multicolumn{2}{|c|}{This paper (LAr TPC)} &  \multicolumn{3}{|c|}{Super-Kamiokande results~\cite{Shiozawa:1998si,Kobayashi:2005pe}}\\
\multicolumn{1}{|c|}{Decay} & Efficiency  & Atmospheric $\nu$&Efficiency & Atmospheric $\nu$ & Published \\
\multicolumn{1}{|c|}{mode} & (\%)  & background  & (\%)  & background & limit \\
\multicolumn{1}{|c|}{} &  &  100~kton$\times$year&  &  92~kton$\times$year &  90\% C.L.\\
\hline\hline
(p1) $p \rightarrow e^+ \; \pi^0$            & 45.3 &   0.1 & 40 & 0.2 & 1.6 $\times 10^{33}$\\
(p2) $p \rightarrow \pi^+ \; \bar{\nu} $     & 41.9 &  78.2 & & &\\
(p3) $p \rightarrow K^+   \; \bar{\nu}$      & 96.8 &   0.1 & 8.6 (prompt-$\gamma$) & 0.7 &2.3 $\times 10^{33}$\\
      &  &    & 6.0 ($K^+ \rightarrow \pi^+\pi^0$) & 0.6 &   \\
(p4) $p \rightarrow \mu^+ \;   \pi^0   $     & 44.8 &   0.8 & 32  & 0.2&\\
(p5) $p \rightarrow \mu^+ \; K^0 $           & 46.7 &   $<0.2$ & 5.4 ($K^0_S\rightarrow \pi^0\pi^0$) & 0.4& \\
           & &    & 7.0 ($K^0_S\rightarrow \pi^+\pi^-$ method 1) & 3.2 & 1.3 $\times 10^{33}$ \\
           & &    & 2.8 ($K^0_S\rightarrow \pi^+\pi^-$ method 2) & 0.3 &\\
(p6) $p \rightarrow e^+ \; K^0 $             & 47.0 &   $<0.2$  & 9.2 ($K^0_S\rightarrow \pi^0\pi^0$) & 1.1& 1.0 $\times 10^{33}$ \\
           & &    & 7.9 ($K^0_S\rightarrow \pi^+\pi^-$ method 1) & 3.6 &\\
           & &    & 1.3 ($K^0_S\rightarrow \pi^+\pi^-$ method 2) & 0.04 &\\
(p7) $p \rightarrow e^+ \; \gamma $          & 98.0 &   $<0.2$ & 73 & 0.1&\\
(p8) $p \rightarrow \mu^+ \; \gamma $        & 98.0 &   $<0.2$ & 51 & 0.2&\\
(p9) $p \rightarrow \mu^- \; \pi^+ \; K^+$   & 97.6 &   0.1 & & &\\
(p10) $p \rightarrow e^+ \; \pi^+ \; \pi^- $  & 18.6 &   2.5 & & &\\
\hline\hline
(n1) $n \rightarrow \pi^0 \; \bar{\nu}$      & 45.1 &  47.4 & & &\\
(n2) $n \rightarrow e^-   \; K^+$            & 96.0 &   $<0.2$ & & &\\
(n3) $n \rightarrow e^+   \;   \pi^-$        & 44.4 &   0.8 & & &\\
(n4) $n \rightarrow \mu^- \;   \pi^+$        & 44.8 &   2.6 & & &\\
\hline
\end{tabular}
\caption{Summary of studied decay modes: signal detection efficiency and expected atmospheric
 neutrino background (normalized to 100~kton$\times$year exposure) after selection cuts.
 Where available, the efficiencies and background results of Super-Kamiokande are given for
 comparison. The published results obtained by Super-Kamiokande are shown
 for completeness~\cite{Shiozawa:1998si,Kobayashi:2005pe}. }
\label{tab:channels}
\end{table*}

The approach to discriminate between signal and background
is based on a set of sequential selection cuts. 
Final state topology and event kinematics provide the selection
criteria. At first, we ignore the cosmic muon-induced background,
and apply cuts until the atmospheric neutrino background
can be considered as irreducible. 

\begin{figure}
\begin{center}
 \epsfysize=8.0cm\epsfxsize=8.0cm
 \epsffile{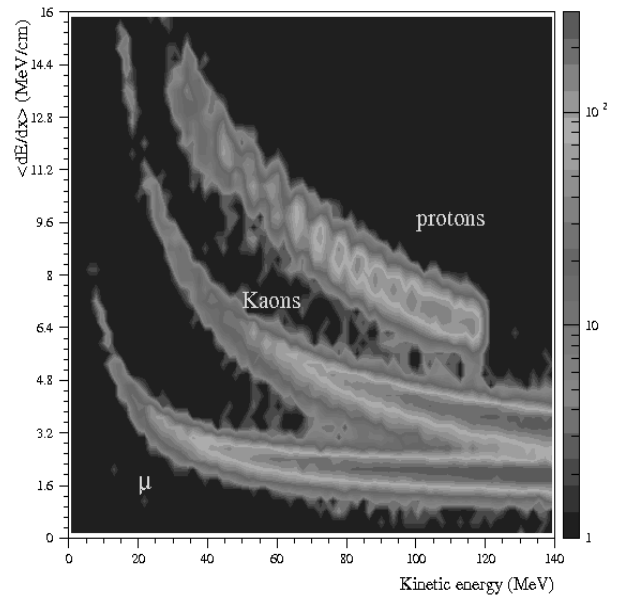}
 \vspace{0.1cm}
 \caption{\small Scatter plot showing the regions populated by fully simulated
 muons, kaons and protons on the $<dE/dx>$ vs. kinetic energy plane.}
 \label{fig:muonkaonproton_scatter}
\end{center}
 \end{figure}

An efficient background reduction demands a good particle identification. 
Results based on a dedicated analysis show that the tagging efficiency
of pions, kaons and protons is above 99\%
with contamination below 1\%. In addition, the muon--pion misidentification is 
around 40\%. The analysis, based on fully-simulated Monte-Carlo events with single
particles, combines several variables in a Multi-layer perceptron
Neural Network architecture: the $\chi^2$ fit to the particle hypothesis on the
$<dE/dx>$ vs kinetic energy plane, the fitted particle mass and the energy released
after the particle decay. Figure~\ref{fig:muonkaonproton_scatter} shows how,
thanks to the fine detector granularity, particles of different species get clearly
separated. Electron and photon identification is based on an algorithm that 
distinguishes single m.i.p. signals versus double m.i.p. signals by using the first 
hits of each identified electromagnetic track.

The list of selection 
cuts to reduce the atmospheric neutrino contamination, final efficiencies and
background estimations are described in sections~\ref{subsec:pdecay}
and~\ref{subsec:ndecay}. 

We then evaluate the backgrounds due to cosmogenic events.
Our strategy is to reduce the contribution 
from this background at the level of the
irreducible atmospheric neutrino background. 
The amount of cosmogenic background
is strongly correlated with the
travel path inside the detector. Thus we 
will cut on the detector fiducial volume, while keeping the
same topology and kinematics criteria used for the neutrino-induced
background (in this way, signal selection efficiencies are
unaltered). The details on the cosmic muon-induced
background estimation are given in Section~\ref{subsec:cosmicbkg}.

\subsection{Proton decay channels}
\label{subsec:pdecay}
 
 The sequential cuts applied for each channel are briefly described
 in the following paragraphs.
The detailed list of cuts for the considered proton decay
 channels are listed in Table~\ref{tab:1}. Survival fraction of 
 signal (first column) and background events from the different atmospheric
 neutrino interactions after 
 selection cuts are applied in succession
 are also listed. Backgrounds are normalized to an exposure of 1 Mton$\times$year.
 The final efficiencies and expected background events after cuts are reported
in Table~\ref{tab:channels}. The published efficiencies, backgrounds and
results obtained by Super-Kamiokande are shown
 for comparison.

\begin{itemize}

\item $p \rightarrow e^+ \; \pi^0 $ channel:
A simulated decay event is shown in Figure~\ref{fig:lar_epi01}.
The two photons (from the $\pi^0$ decay) and the positron flying in the opposite
direction are clearly visible. In the chosen readout view, the event spreads over about
120$\times$100~cm$^2$. 
Figure~\ref{fig:pdk_e+pi0_kine} shows the distributions of the following
reconstructed kinematical quantities: the electron
momentum, the total momentum imbalance, the invariant mass and
the total energy. The distributions are split into the case where
the pion leaves the nucleus (full histograms) and the case where it is
absorbed (empty histograms). 
The electron momentum histrogram has an arrow placed at 460~MeV/c,
the expected value without Fermi motion and detector effects.
It seems clear that two different set of cuts could be implemented
to optimize the signal over background ratio in both cases.
However, an attempt to look for ``inclusive'' decays $p\rightarrow e^+(\pi^0)$
without condition on the $\pi^0$, yields order of magnitudes worse
background conditions, and was not considered further.

\begin{figure}
\centering
 \setlength{\unitlength}{1mm}
 \begin{picture}(110,80)
   \put(18,0){\includegraphics[width=55\unitlength]{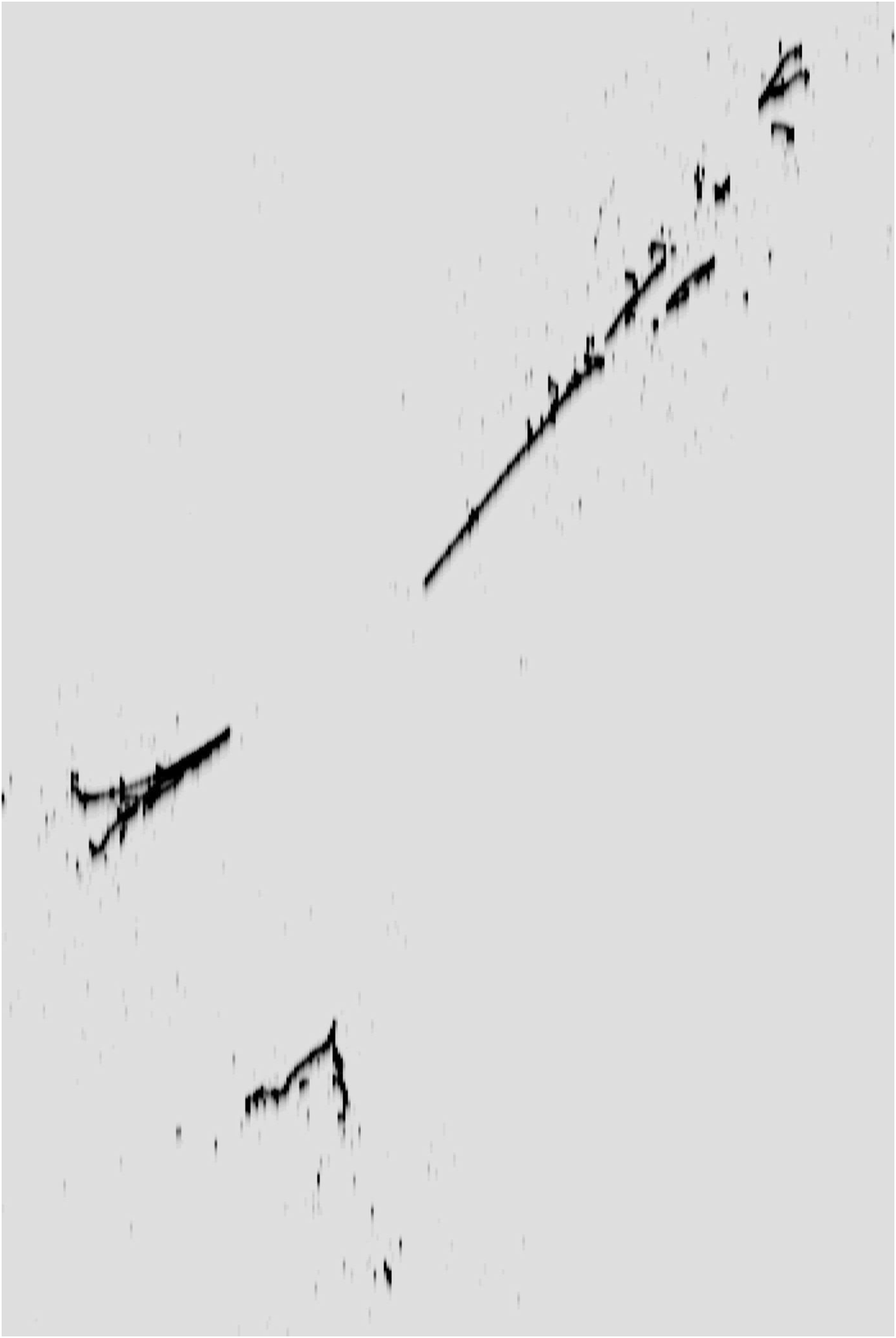}}
      \put(18,0){\vector(1,0){55}}
   \put(40,4){\makebox(0,0)[tl]{Wire number (120~cm)}}
   \put(18,0){\vector(0,1){75}}
   \put(13,50){\rotatebox{90}{$t_{drift}$ (100~cm)}}
   \put(50,50){\makebox(0,0)[tl]{$e^+$}}
   \put(25,40){\makebox(0,0)[tl]{$\gamma$}}
   \put(40,20){\makebox(0,0)[tl]{$\gamma$}}
 \end{picture}
\caption{Simulated $p \rightarrow e^{+}\pi^0$ event.  The 
 displayed area covers $120\times 100$~cm$^2$.}
\label{fig:lar_epi01}
\end{figure}

 \begin{figure}
\begin{center}
 \epsfysize=9.5cm\epsfxsize=9.5cm
 \hspace*{0.0cm}\epsffile{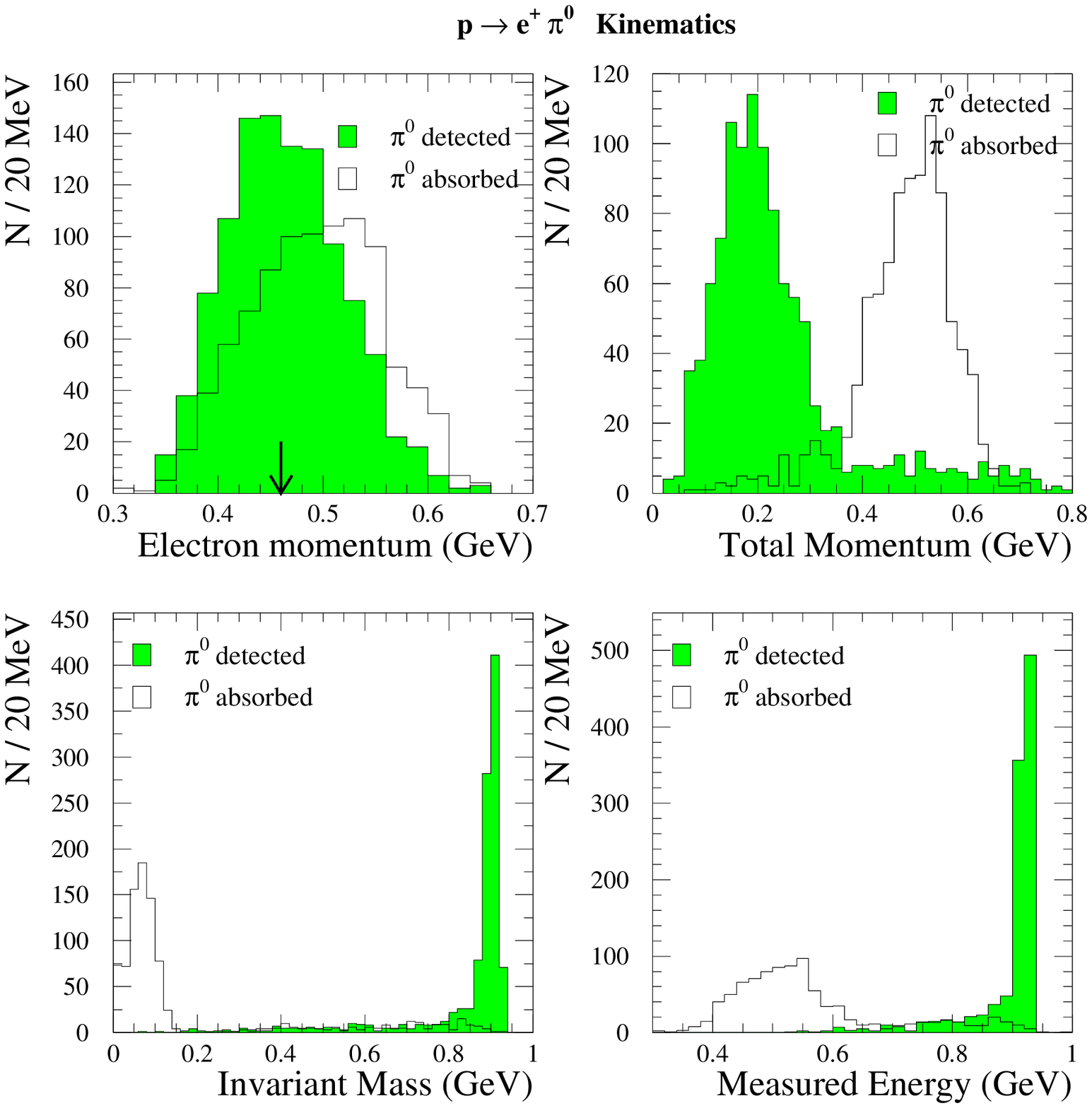}
 \caption{\small $p \rightarrow e^+ \pi^0$ channel:
Distributions of some kinematic variables for the exclusive
(full histograms) and the inclusive (empty histograms) scenarios.
The arrow in the first plot indicates the value that would have
the positron momentum if no Fermi motion and no detector effects
were present.}
 \label{fig:pdk_e+pi0_kine}
\end{center}
\end{figure}

The list of cuts is presented in Table~\ref{tab:1}.
The idea is to have a balanced event, with all particles
identified as such, and with a total visible energy close to the proton mass
(see Figure~\ref{fig:pdk_e+pi0_minv}).
Only one background event for 1~Mton$\times$year exposure survives the cuts,
for a signal efficiency of about 45\%. 

 \begin{figure}
\begin{center}
 \epsfysize=8.0cm\epsfxsize=8.0cm
 \hspace*{0.1cm}\epsffile{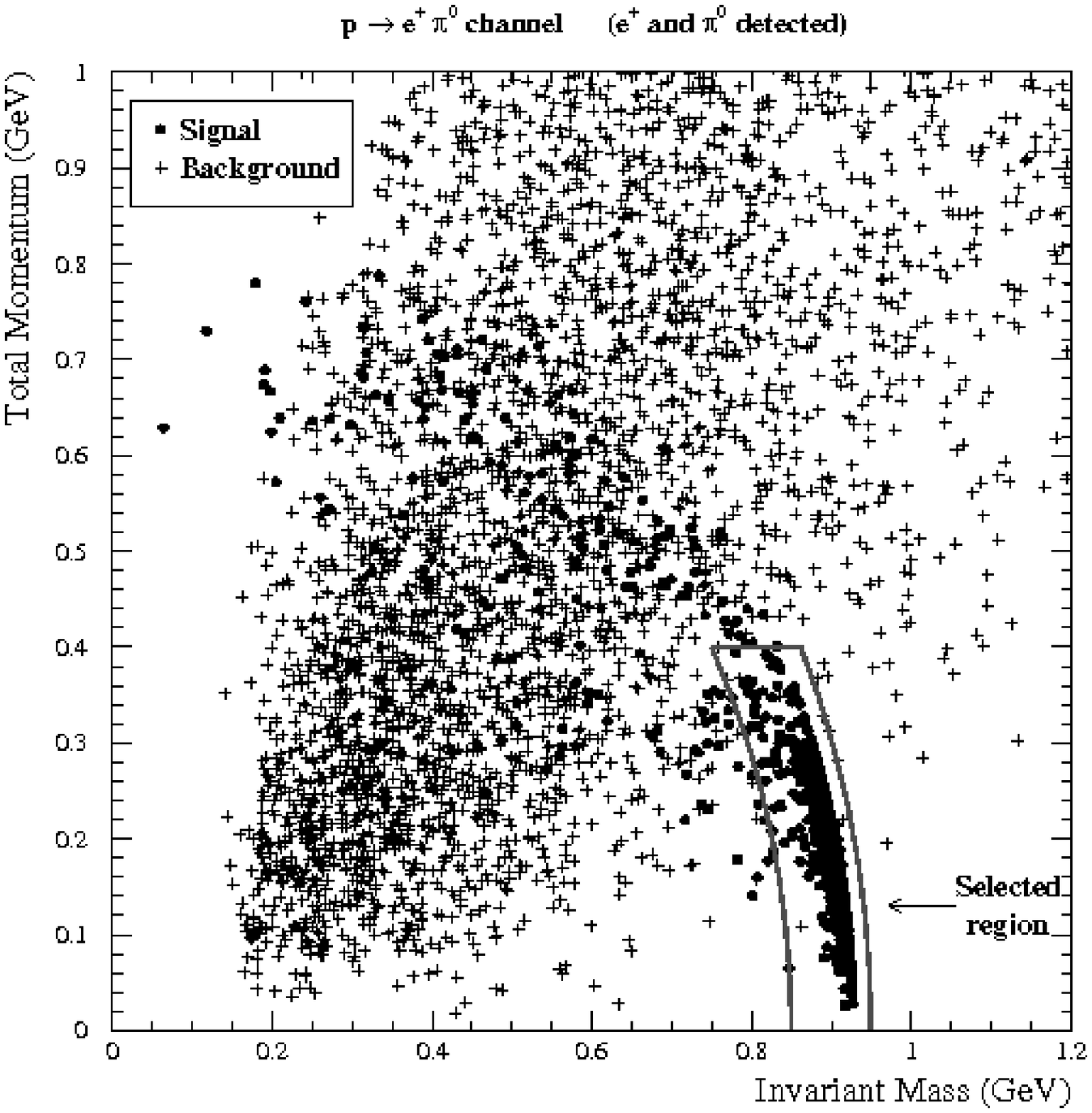}
 \caption{\small Kinematic cut in the $p \rightarrow e^+ \pi^0$ channel:
  in the plane defined by the invariant mass and the total momentum,
  crosses represent background and spots signal events.
  The band indicates the cut region (0.86~GeV~$<$~Total Energy~$<$~0.95 GeV),
  i.e. all events inside the band are accepted.}
 \label{fig:pdk_e+pi0_minv}
\end{center}
\end{figure}

{\small 
\begin{table*}
\begin{center}
\begin{tabular}{|c|c|c|c|c|c|c|c|}
\cline{2-8}
\multicolumn{1}{c|}{\bf{}} & Efficiency (\%) & \multicolumn{6}{c|}{\bf{Atmospheric neutrino sources}} \\
\cline{2-8}
\multicolumn{1}{c|}{\bf{Cuts}} & \bf{(p1) $p \rightarrow e^+   \;   \pi^0$} &
\bf{$\nu_e$ CC}& \bf{$\bar{\nu}_e$ CC} & \bf{$\nu_{\mu}$ CC}&
\bf{$\bar{\nu}_{\mu}$ CC} & \bf{$\nu$ NC} & \bf{$\bar{\nu}$ NC} \\
\hline
One $\pi^0$                         & 54.0\% & 6604 & 2135 & 15259 & 5794 & 8095 & 3103 \\\hline
One e-shower + no other charged tracks                          & 50.9\% & 1188 &  656 &     1 &    0 &    0 &    0 \\\hline
p$_{tot} <$ 0.4 GeV                 & 46.7\% &  454 &  127 &     0 &    0 &    0 &    0 \\\hline
0.86 GeV $<$ E$_{vis}$ $<$ 0.95 GeV & 45.3\% &    1 &    0 &     0 &    0 &    0 &    0 \\\hline
\hline

\hline
\multicolumn{1}{c|}{\bf{Cuts}} & \bf{(p2) $p \rightarrow \pi^+ \; \bar{\nu}             $} &
\bf{$\nu_e$ CC}& \bf{$\bar{\nu}_e$ CC} & \bf{$\nu_{\mu}$ CC}&
\bf{$\bar{\nu}_{\mu}$ CC} & \bf{$\nu$ NC} & \bf{$\bar{\nu}$ NC} \\
\hline
No e-shower, no muons, no $\pi^0$                          &  92.6\% &     0 &     0 &    34 &     0 & 56515 & 26482 \\ \hline
One charged pion                    &  55.7\% &     0 &     0 &     8 &     0 &  5632 &  2027 \\ \hline
No  protons                         &  50.0\% &     0 &     0 &     4 &     0 &  2930 &  1136 \\ \hline
0.35 GeV $<$ Total E $<$ 0.65 GeV   &  41.9\% &     0 &     0 &     2 &     0 &   605 &   175 \\ \hline
\hline

\hline
\multicolumn{1}{c|}{\bf{Cuts}} & \bf{(p3) $p \rightarrow K^+   \; \bar{\nu} $} &
\bf{$\nu_e$ CC}& \bf{$\bar{\nu}_e$ CC} & \bf{$\nu_{\mu}$ CC}&
\bf{$\bar{\nu}_{\mu}$ CC} & \bf{$\nu$ NC} & \bf{$\bar{\nu}$ NC} \\
\hline
One kaon                 &  96.8\% &  308 &  36 &  871 & 146 &  282 &  77 \\ \hline
No other charged tracks, no $\pi^0$   &  96.8\% &    0 &   0 &    0 &   0 &   57 &   9 \\ \hline
E$_{vis}$ $<$ 0.8 GeV    &  96.8\% &    0 &   0 &    0 &   0 &    1 &   0 \\ \hline
\hline

\hline
\multicolumn{1}{c|}{\bf{Cuts}} & \bf{(p4) $p \rightarrow \mu^+ \;   \pi^0               $} &
\bf{$\nu_e$ CC}& \bf{$\bar{\nu}_e$ CC} & \bf{$\nu_{\mu}$ CC}&
\bf{$\bar{\nu}_{\mu}$ CC} & \bf{$\nu$ NC} & \bf{$\bar{\nu}$ NC} \\
\hline
One muon, one $\pi^0$                            &  52.8\% &     0 &     0 & 11334 &  4452 &     0 &     0 \\ \hline
No protons, no charged pions                    &  50.0\% &     0 &     0 &  1754 &  1369 &     0 &     0 \\ \hline
0.86 GeV $<$ Total E $<$ 0.93 GeV   &  45.3\% &     0 &     0 &    64 &    41 &     0 &     0 \\ \hline
Total Momentum $< 0.5$ GeV          &  44.8\% &     0 &     0 &     5 &     3 &     0 &     0 \\ \hline
\hline

\hline
\multicolumn{1}{c|}{\bf{Cuts}} & \bf{(p5) $p \rightarrow \mu^+   \; K^0_S $} &
\bf{$\nu_e$ CC}& \bf{$\bar{\nu}_e$ CC} & \bf{$\nu_{\mu}$ CC}&
\bf{$\bar{\nu}_{\mu}$ CC} & \bf{$\nu$ NC} & \bf{$\bar{\nu}$ NC} \\
\hline
One muon  + 2 charged or 2 neutral pions               
                         &  100\% & 8178 &  2771 &  106861 & 27274 & 7099 & 2540 \\ \hline
0.4 $<$ Invariant mass of pions $<$ 0.6 GeV          &  97\%  & 0 & 0 &  5   &   8 & 6 & 2 \\ \hline
$p_{tot} <$ 0.6 GeV      &  93.4\%  &     0 &     0 &      0 &     0 &     0 &     0 \\ \hline
\hline

\hline
\multicolumn{1}{c|}{\bf{Cuts}} & \bf{(p6) $p \rightarrow e^+   \; K^0_S $} &
\bf{$\nu_e$ CC}& \bf{$\bar{\nu}_e$ CC} & \bf{$\nu_{\mu}$ CC}&
\bf{$\bar{\nu}_{\mu}$ CC} & \bf{$\nu$ NC} & \bf{$\bar{\nu}$ NC} \\
\hline
One e-shower + 2 charged or 2 neutral pions                  
                         &  100\% & 59759 &  11673 &  31 & 0 &  2 &  1 \\ \hline
0.4 $<$ Invariant mass of pions $<$ 0.6 GeV         &  97.0\% &  2 &  2 &  0 & 0 &  0 &  0 \\ \hline
$p_{tot}<$ 0.6 GeV   &  94.0\% &    0 &   0 &   0 & 0 &  0 &   0 \\ \hline
\hline

\hline
\multicolumn{1}{c|}{\bf{Cuts}} & \bf{(p7) $p \rightarrow e^+   \; \gamma $} &
\bf{$\nu_e$ CC}& \bf{$\bar{\nu}_e$ CC} & \bf{$\nu_{\mu}$ CC}&
\bf{$\bar{\nu}_{\mu}$ CC} & \bf{$\nu$ NC} & \bf{$\bar{\nu}$ NC} \\
\hline
One e-shower + no other charged track                 
                         &  100\% &  32434 &  6837 &  0 & 0 &  0 &  0 \\ \hline
Only one photon          &  99.0\% &  110 &  11 &  0 & 0 &  0 &  0 \\ \hline
p$_\gamma$ $>$ 0.2 GeV   &  98.0\% &    0 &   0 &   0 & 0 &  0 &   0 \\ \hline
\hline

\hline
\multicolumn{1}{c|}{\bf{Cuts}} & \bf{(p8) $p \rightarrow \mu^+   \; \gamma $} &
\bf{$\nu_e$ CC}& \bf{$\bar{\nu}_e$ CC} & \bf{$\nu_{\mu}$ CC}&
\bf{$\bar{\nu}_{\mu}$ CC} & \bf{$\nu$ NC} & \bf{$\bar{\nu}$ NC} \\
\hline
One muon + no other charged track                 
                         &  100\% &  5302 &  1878 &  54889 & 15872 &  4680 &  1764 \\ \hline
Only one photon          &  99.0\%  &     7 &     4 &    164 &    13 &     9 &     7 \\ \hline
p$_\gamma$ $>$ 0.2 GeV   &  98.0\%  &     0 &     0 &      0 &     0 &     0 &     0 \\ \hline
\hline

\hline
\multicolumn{1}{c|}{\bf{Cuts}} & \bf{(p9) $p \rightarrow \mu^- \; \pi^+ \; K^+          $} &
\bf{$\nu_e$ CC}& \bf{$\bar{\nu}_e$ CC} & \bf{$\nu_{\mu}$ CC}&
\bf{$\bar{\nu}_{\mu}$ CC} & \bf{$\nu$ NC} & \bf{$\bar{\nu}$ NC} \\
\hline
One Kaon                            &  98.8\% &   308 &    36 &   871 &   146 &   282 &    77 \\ \hline
One muon                            &  98.2\% &     1 &     0 &   867 &   146 &     0 &     0 \\ \hline
No  e-showers                       &  98.2\% &     0 &     0 &   844 &   145 &     0 &     0 \\ \hline
0.6 GeV $<$ Total E $<$ 1 GeV       &  97.6\% &     0 &     0 &     1 &     0 &     0 &     0 \\ \hline
\hline

\hline
\multicolumn{1}{c|}{\bf{Cuts}} & \bf{(p10) $p \rightarrow e^+   \;   \pi^+ \;   \pi^-    $} &
\bf{$\nu_e$ CC}& \bf{$\bar{\nu}_e$ CC} & \bf{$\nu_{\mu}$ CC}&
\bf{$\bar{\nu}_{\mu}$ CC} & \bf{$\nu$ NC} & \bf{$\bar{\nu}$ NC} \\
\hline
One e-shower, no  muons                           & 100\% & 59755 & 11673 &     0 &     0 &     0 &     0 \\ \hline
Two charged pions, no  protons                         &  19.4\% &   714 &   302 &     0 &     0 &     0 &     0 \\ \hline
0.65 GeV $<$ Total E $<$ 1 GeV      &  19.0\% &    33 &     8 &     0 &     0 &     0 &     0 \\ \hline
Total Momentum $<$ 0.57 GeV         &  18.6\% &    21 &     4 &     0 &     0 &     0 &     0 \\ \hline
\hline

\end{tabular}
\vspace{0.1cm}
\caption{\small Detailed list of cuts for the considered proton decay
 channels. Survival fraction of 
 signal (first column) and background events through event 
 selections applied in succession. Backgrounds are normalized to an exposure of 1 Mton$\times$year.}
\label{tab:1}
\end{center}
\end{table*}
}

\item $p \rightarrow \pi^+ \; \bar{\nu}$ channel:
  Almost 45\% of the events that belong to this channel can not be
detected since the $\pi^+$ gets absorbed by the nucleus.
The cuts are based on the requirement of absence of charged leptons,
protons, neutral pions, the presence of one charged pion and
a total energy between 350 and 650~MeV. The result is that, for a
$\sim$42\% efficiency, the expected background at 1~kton$\times$year
exposure is $\sim$0.8~events.

\item $p \rightarrow K^+ \; \bar{\nu}$ channel:
 this is a quite clean
channel due to the presence of a strange meson and no other particle in
the final state (see Figure~\ref{fig:lar_nuk}). The kaon particle identification
is performed and applying the cuts
listed in Table~\ref{tab:1} yields an
efficiency $\sim$97\% for a negligible background. The correlation
of the reconstructed
invariant mass and total momentum is shown in Figure~\ref{fig:pdk_k+anu_minv}.

\begin{figure}
\centering
 \setlength{\unitlength}{1mm}
 \begin{picture}(110,80)
   \put(18,0){\includegraphics[width=55\unitlength]{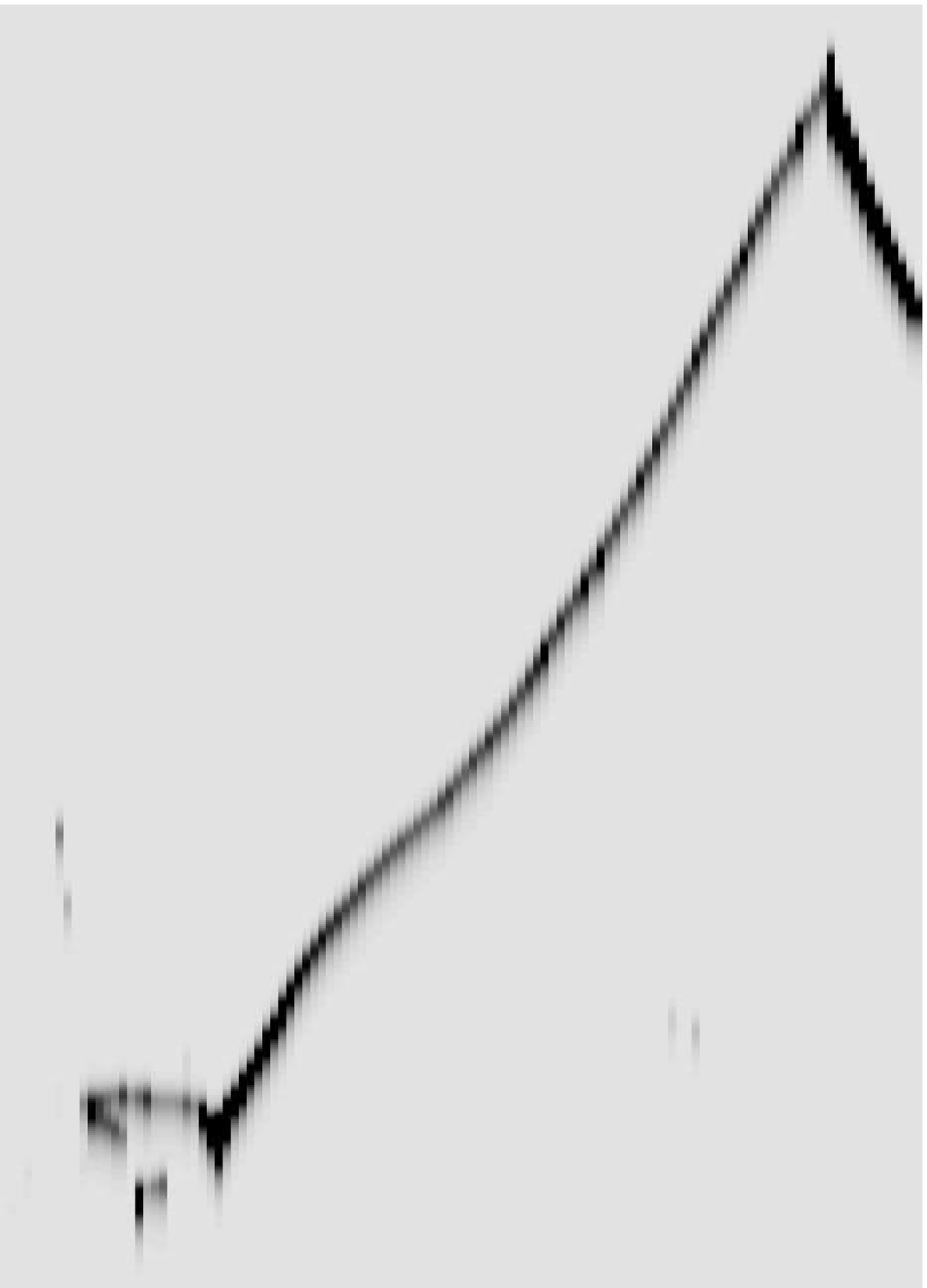}}
      \put(18,0){\vector(1,0){55}}
   \put(40,4){\makebox(0,0)[tl]{Wire number (34~cm)}}
   \put(18,0){\vector(0,1){75}}
   \put(14,50){\rotatebox{90}{$t_{drift}$ (90~cm)}}
   \put(65,60){\makebox(0,0)[tl]{$K^+$}}
   \put(40,40){\makebox(0,0)[tl]{$\mu^+$}}
   \put(25,20){\makebox(0,0)[tl]{$e^+$}}
 \end{picture}
 \caption{Simulated $p \rightarrow K^+ \; \bar{\nu}$ event. The 
 displayed area covers $34\times 90$~cm$^2$.}
\label{fig:lar_nuk}
\end{figure}

 \begin{figure}
\begin{center}
 \epsfysize=8.0cm\epsfxsize=8.0cm
 \hspace*{0.1cm}\epsffile{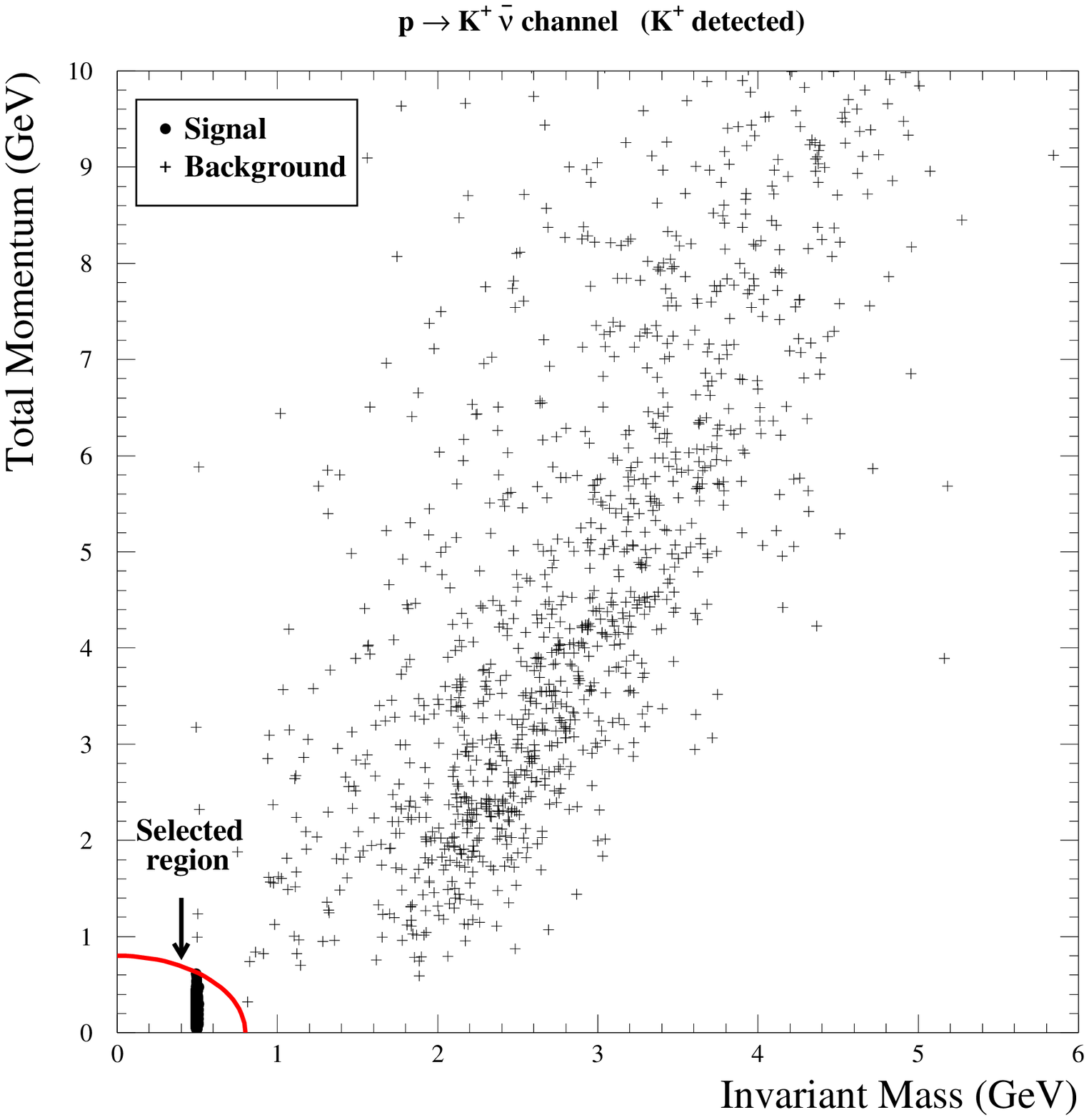}
 \caption{\small Kinematic cut in the $p \rightarrow K^+ \bar{\nu}$ channel:
  in the plane defined by the invariant mass and the total momentum,
  crosses represent background and spots signal events.
  The band indicates the cut region (Total Energy $<$ 0.8 GeV),
  i.e. all events in the band are accepted.}
 \label{fig:pdk_k+anu_minv}
\end{center}
\end{figure}

\item $p \rightarrow \mu^+ \; \pi^0$ channel:
Almost 53\% of the times the $\pi^0$ is detected. In this case, cuts are similar
to the $e^+ \pi^0$ channel. The efficiency remains high ($\sim$45\%), while the
background is $\simeq 8$~events for 1~Mton$\times$year.

\item $p \rightarrow e^+ \; K^0$ and  $p \rightarrow \mu^+ \; K^0$ channels: 
We concentrate  on final states having a K$^0_S$, since a large
fraction of the K$^0_L$ will leave the detector without decaying or will
suffer hadronic interactions. In Figure~\ref{fig:ek0} simulated events of 
$p \rightarrow e^+ + K^0_S$ 
and $p \rightarrow \mu^+ + K^0_S$ decays
are shown, where the K$^0_S$ 
decays into charged pions. A $p \rightarrow \mu^+ + K^0$ decay where
K$^0$ decays into neutral pions is shown in
Figure~\ref{fig:muk0_2}. 
A sophisticated treatment to recover 
K$^0_L$ events has been neglected at this stage. K$^0_S$'s mainly decay to two 
pions (either charged or neutral). Simple requirements (an
identified lepton in the final state accompanied by two charged
(neutral) pions, invariant mass of the pion
system consistent with the K$^0$ mass and total momentum below 0.6~GeV) 
result in a negligible background contamination for 
signal selection efficiencies above 90$\%$ (see Table~\ref{tab:1}). 
The overall efficiency for the $p \rightarrow \mu^+ (e^+) \; K^0$ 
channel is therefore $46.7\%$ ($47\%$).

\begin{figure*}
\begin{center}
\includegraphics[width=0.25\textwidth]{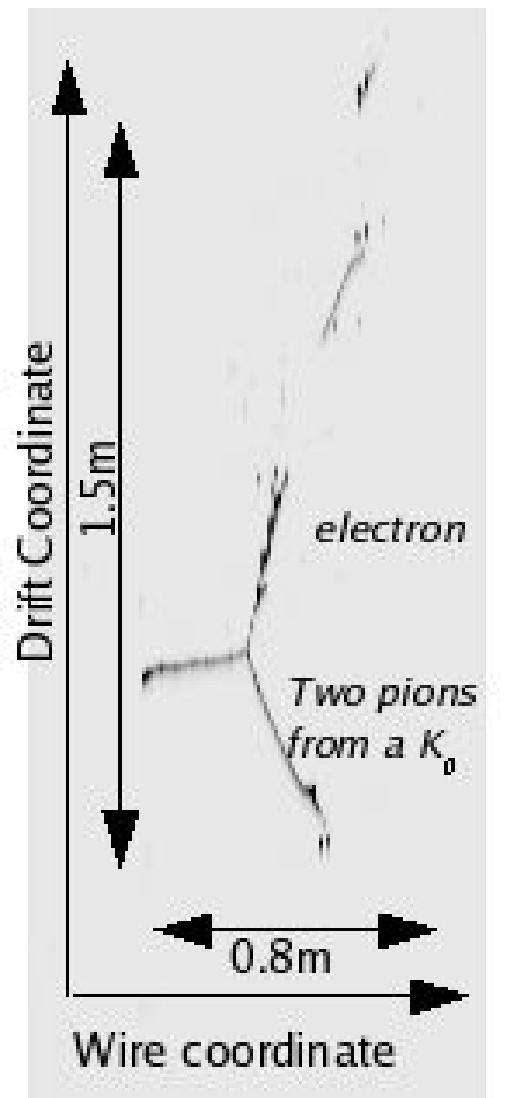}
\includegraphics[width=0.4\textwidth]{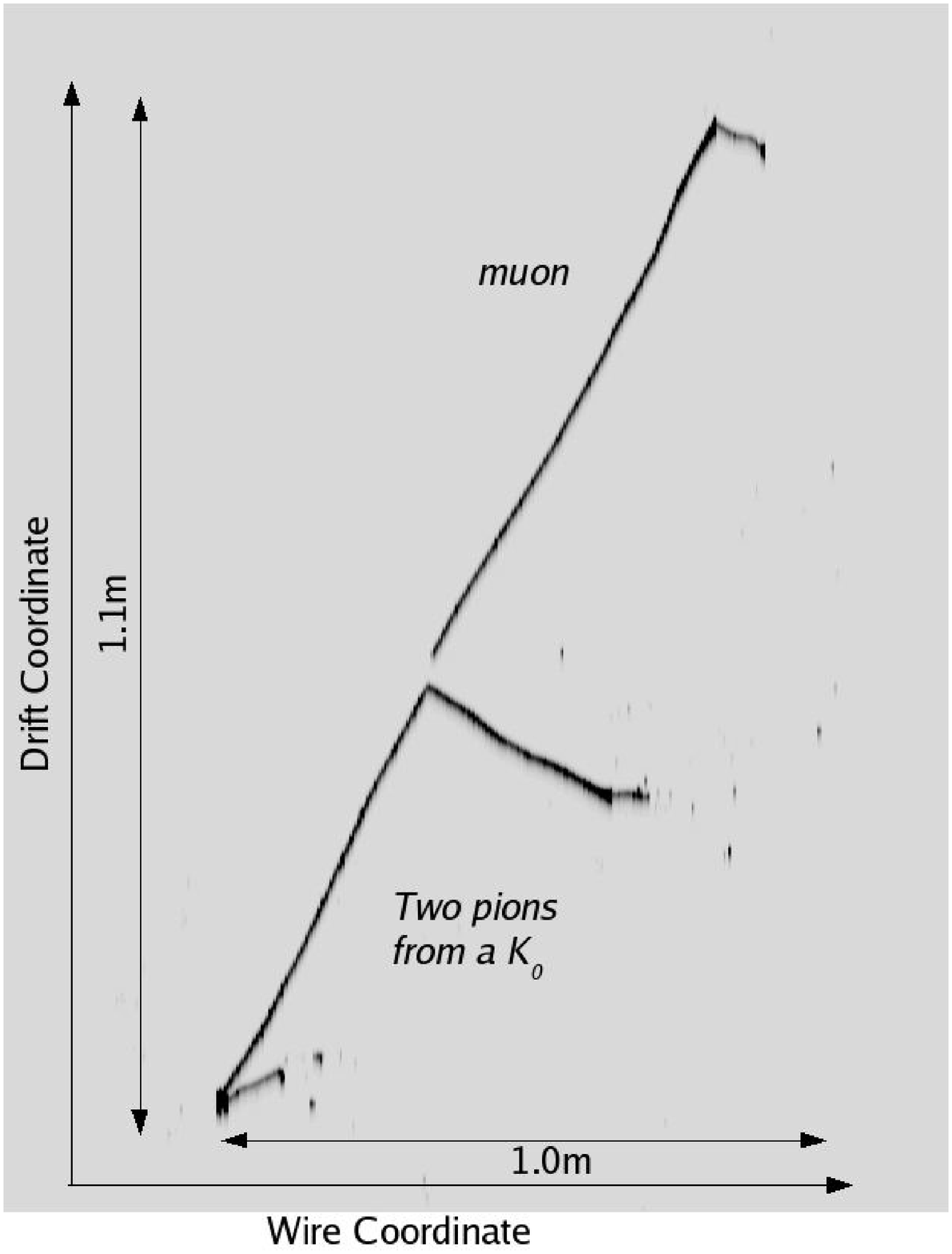}
 \caption{\small Simulated $p \rightarrow e^+ + K^0$ (left)
 and $p \rightarrow \mu^+ + K^0$ (right) decays. The neutral kaons decay into two charged pions.}
 \label{fig:ek0}
\end{center}
\end{figure*}

\begin{figure*}
\begin{center}
\includegraphics[width=0.6\textwidth]{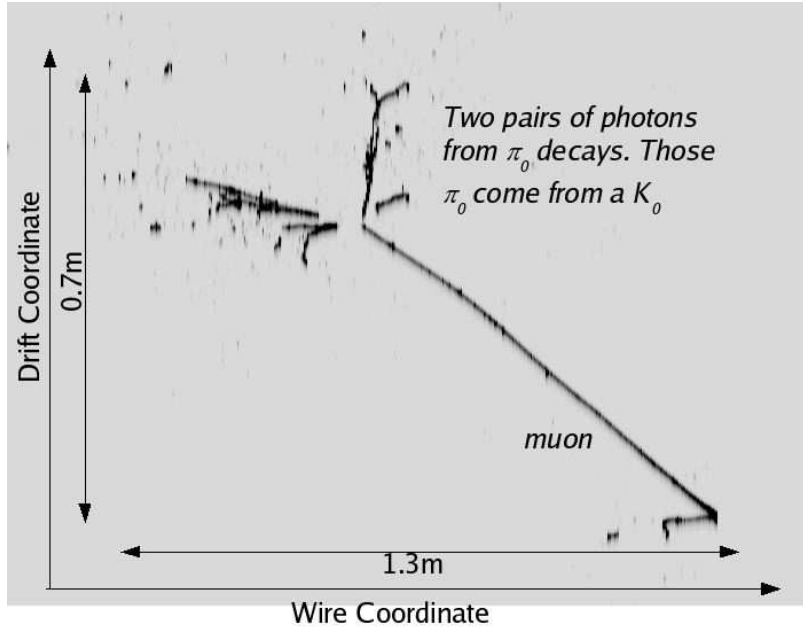}
 \caption{\small Simulated $p \rightarrow \mu^+ + K^0$ decay. The neutral kaon decays into two neutral pions.}
 \label{fig:muk0_2}
\end{center}
\end{figure*}

\item $p \rightarrow e^+ \gamma$ and $p \rightarrow \mu^+ \gamma$ channels: 
These channels provide a very clean signal thanks to efficient electron and photon 
separation (see e.g. Ref.~\cite{Meregaglia:2006du} for a discussion in the context of $e/\pi^0$ separation). 
The simple final event topology 
(a single charged lepton accompanied by an energetic photon) 
allows to reduce the expected
background to a negligible level while keeping a signal selection
efficiency close to 100$\%$ (see Table~\ref{tab:1}).
Simulated events of $p \rightarrow e^+ + \gamma$ decay and 
$p \rightarrow \mu^+ + \gamma$ are shown in Figure~\ref{fig:egamma}. 

\begin{figure}
\begin{center}
\includegraphics[height=10cm]{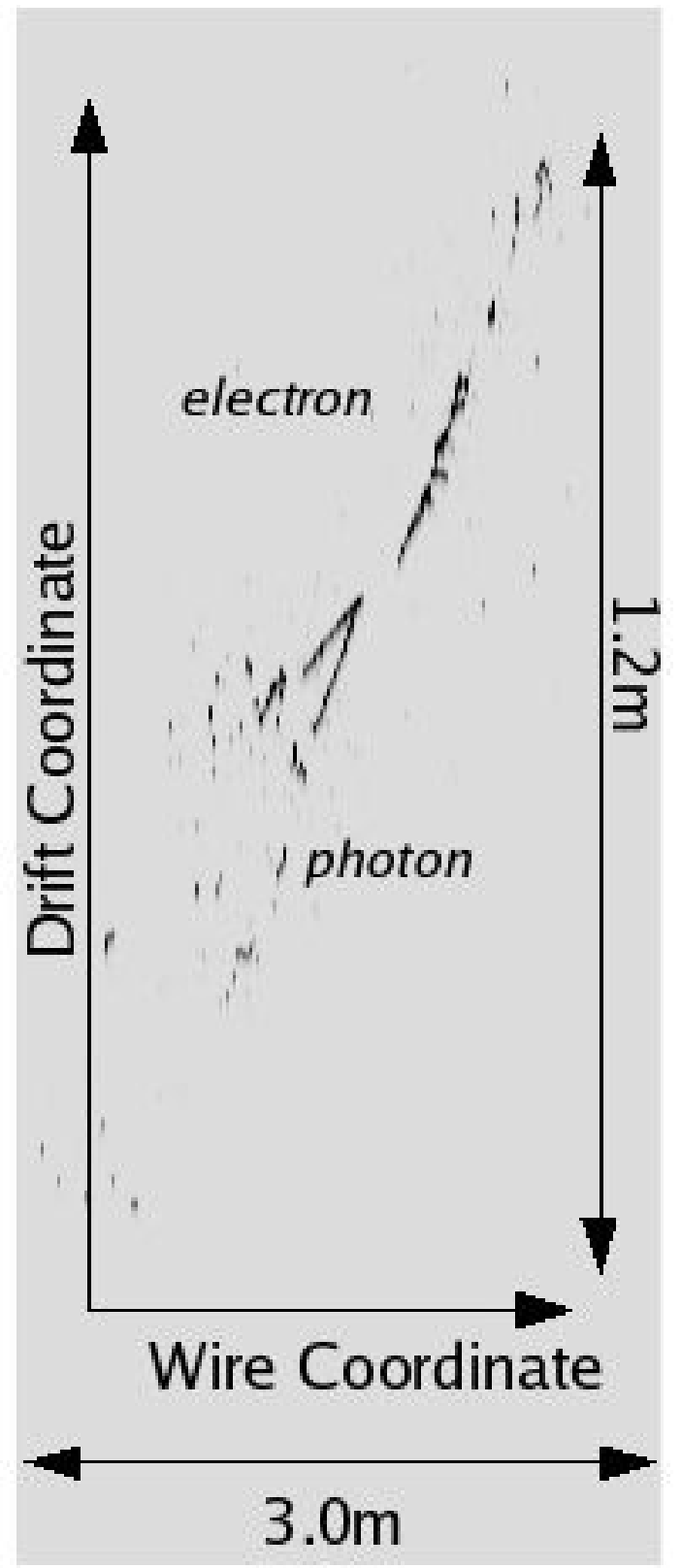}
\includegraphics[height=10cm]{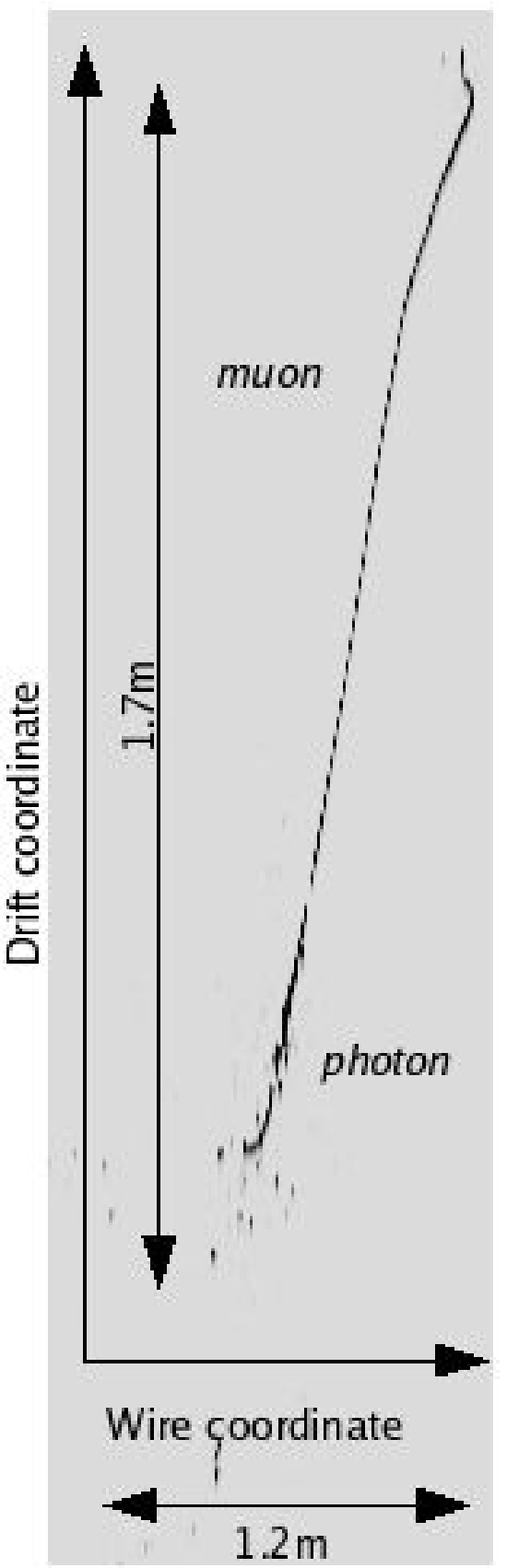}
 \caption{\small Simulated $p \rightarrow e^+ + \gamma$ 
 and $p \rightarrow \mu^+ + \gamma$ decays.}
 \label{fig:egamma}
\end{center}
\end{figure}

\item $p \rightarrow \mu^- \; \pi^+ \; K^+$ channel:
by tagging the presence of 
a $K^+$ and a $\mu^-$, the background
is reduced at the level of $\sim$1 event for 1~Mton$\times$year exposure.
A final cut on the total visible energy removes any background
for a signal of $\simeq$~97\%.

\item $p \rightarrow e^+ \; \pi^+ \; \pi^-$ channel:
The most favorable scenario occurs when the three particles are detected.
A tight cut on the total visible energy (0.65$<$$E_{vis}$$<$1~GeV) complemented with
a cut on the total momentum ($p_{tot}$$<$570~MeV/c)
are sufficient to reduce the contamination of 25 events for a
1~Mton $\times$year exposure. On the other hand, the cuts remove less than
1\% of the signal events.

\end{itemize}

\subsection{Neutron decay channels}
\label{subsec:ndecay}

The sequential cuts applied for each channel are briefly described
 in the following paragraphs.
The detailed list of cuts for the considered neutron decay
 channels are listed in Table~\ref{tab:14}. Survival fraction of 
 signal (first column) and background events from the different atmospheric
 neutrino interactions after 
 selection cuts are applied in succession
 are also listed. Backgrounds are normalized to an exposure of 1 Mton$\times$year.
 The final efficiencies and expected background events after cuts are reported
in Table~\ref{tab:channels}.

\begin{figure}
\begin{center}
 \epsfysize=8.0cm\epsfxsize=8.0cm
 \hspace*{0.1cm}\epsffile{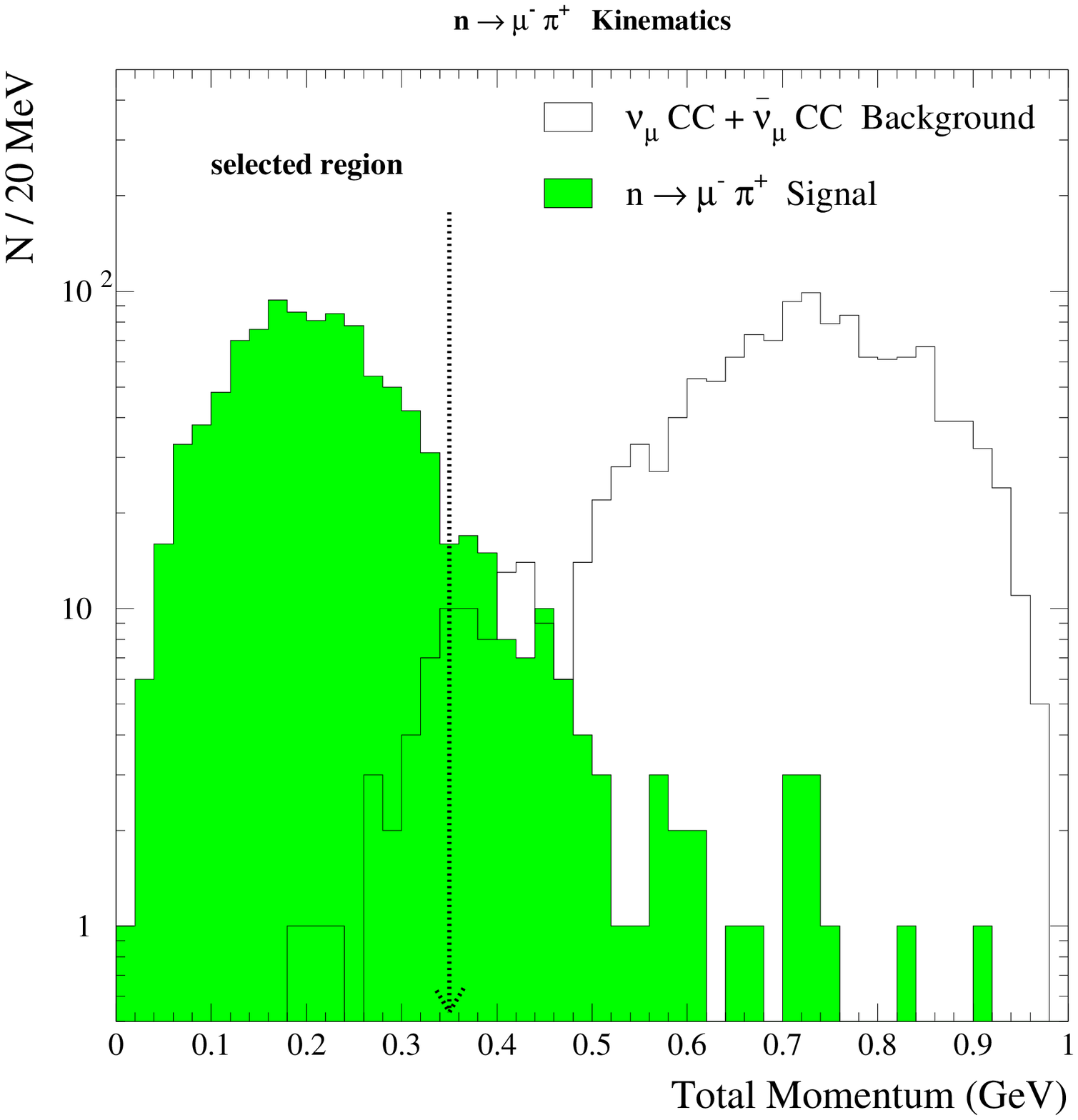}
 \caption{\small $n \rightarrow \mu^- \; \pi^+$ channel:
  distribution of total momentum for events surviving the sixth cut
  of Table~\ref{tab:14}.}
 \label{fig:ndk_mu-pi+_totp_bkg}
\end{center}
\end{figure}

\begin{itemize}

\item $n \rightarrow \pi^0 \; \bar{\nu}$ channel:
About $\sim$45\% of the signal events are ``invisible'' because the $\pi^0$
is absorbed in the nucleus. The rest can be disentangled from the background by
cutting on the total visible energy and on the total momentum ($>$0.35~GeV).
The final efficiency is $\sim$45\% for $\sim$0.5 background events
at 1~kton$\times$year exposure.

\item $n \rightarrow e^- \; K^+$ channel:
We profit here from the presence of one kaon, one electron and the absence
of muons and pions in the final state. These requirements, together with
a loose cut on the visible energy (0.75$<$$E_{vis}$$<$0.95~GeV), eliminates the
background for an almost untouched efficiency.

\item $n \rightarrow e^+ \; \pi^-$ channel:
This channel is similar to the $p \rightarrow e^+ \; \pi^0$ previously reported.
When the $\pi^-$ is detected, a good efficiency can be reached for a negligible
background by applying two simple cuts to bound the total energy
(0.75$<$$E_{vis}$$<$1~GeV) and the positron momentum (0.35$<$$p_{e}$$<$0.6~GeV).

\item $n \rightarrow \mu^- \; \pi^+$ channel:
  This channel is treated in a similar way to the $n \rightarrow e^+ \; \pi^-$.
  The distribution of the total momentum for signal and background
events is shown 
in Figure~\ref{fig:ndk_mu-pi+_totp_bkg} and the cuts in Table~\ref{tab:14}.
We require the presence of a pion on the final state and a cut on the total energy.
The plot shows the position of the last cut on total momentum.
  The final efficiency is $\sim$45\% for the exclusive 
 channel and almost no background events 
expected.

\end{itemize}

{\small 
\begin{table*}
\begin{center}
\begin{tabular}{|c|c|c|c|c|c|c|c|}\cline{2-8}
\multicolumn{1}{c|}{\bf{}} & Efficiency (\%) & \multicolumn{6}{c|}{\bf{Atmospheric neutrino sources}} \\
\cline{2-8}

\multicolumn{1}{c|}{\bf{Cuts}} & \bf{(n1) $n \rightarrow \pi^0 \; \bar{\nu}             $} &
\bf{$\nu_e$ CC}& \bf{$\bar{\nu}_e$ CC} & \bf{$\nu_{\mu}$ CC}&
\bf{$\bar{\nu}_{\mu}$ CC} & \bf{$\nu$ NC} & \bf{$\bar{\nu}$ NC} \\
\hline
One  $\pi^0$                        &  56.2\% &  6604 &  2135 & 15259 &  5794 &  8095 &  3103 \\ \hline
No muons, no electrons, no charged pions                    &  56.1\% &     0 &     0 &     2 &     0 &  4722 &  1840 \\ \hline
No protons                          &  52.6\% &     0 &     0 &     0 &     0 &  2964 &  1184 \\ \hline
0.35 GeV $<$ Total E   $<$ 0.55 GeV &  45.4\% &     0 &     0 &     0 &     0 &   469 &   181 \\ \hline
Total Momentum   $>$ 0.35 GeV       &  45.1\% &     0 &     0 &     0 &     0 &   362 &   112 \\ \hline

\hline
\multicolumn{1}{c|}{\bf{Cuts}} & \bf{(n2) $n \rightarrow e^-   \; K^+                   $} &
\bf{$\nu_e$ CC}& \bf{$\bar{\nu}_e$ CC} & \bf{$\nu_{\mu}$ CC}&
\bf{$\bar{\nu}_{\mu}$ CC} & \bf{$\nu$ NC} & \bf{$\bar{\nu}$ NC} \\
\hline
One e-shower, one kaon                        &  97.0\% &   299 &    36 &    11 &     0 &     0 &     0 \\ \hline
No  $\pi^0$ , no  muons                           &  97.0\% &   138 &    14 &     0 &     0 &     0 &     0 \\ \hline
No  charged pions                   &  97.0\% &    80 &     5 &     0 &     0 &     0 &     0 \\ \hline
0.75 GeV $<$ Total E $<$ 0.95 GeV   &  96.0\% &     0 &     0 &     0 &     0 &     0 &     0 \\ \hline

\hline
\multicolumn{1}{c|}{\bf{Cuts}} & \bf{(n3) $n \rightarrow e^+   \;   \pi^-               $} &
\bf{$\nu_e$ CC}& \bf{$\bar{\nu}_e$ CC} & \bf{$\nu_{\mu}$ CC}&
\bf{$\bar{\nu}_{\mu}$ CC} & \bf{$\nu$ NC} & \bf{$\bar{\nu}$ NC} \\
\hline
One e-shower, one charged pion                    &  59.6\% &  8137 &  2755 &     6 &     0 &     0 &     0 \\ \hline
No  $\pi^0$ , no  muons, no  protons                         &  57.4\% &  3855 &  1282 &     0 &     0 &     0 &     0 \\ \hline
0.75 GeV $<$ Total E $<$ 1 GeV      &  52.4\% &   499 &   187 &     0 &     0 &     0 &     0 \\ \hline
0.35 GeV $< P_{positron} <$ 0.6 GeV &  51.3\% &   216 &    73 &     0 &     0 &     0 &     0 \\ \hline
Total Momentum $<$ 0.35 GeV         &  44.4\% &     7 &     1 &     0 &     0 &     0 &     0 \\ \hline

\hline
\multicolumn{1}{c|}{\bf{Cuts}} & \bf{(n4) $n \rightarrow \mu^- \;   \pi^+  $} &
\bf{$\nu_e$ CC}& \bf{$\bar{\nu}_e$ CC} & \bf{$\nu_{\mu}$ CC}&
\bf{$\bar{\nu}_{\mu}$ CC} & \bf{$\nu$ NC} & \bf{$\bar{\nu}$ NC} \\
\hline
One muon, one charged pion        &  59.4\% &  1559 & 454 & 15931 &  6569 & 2291 & 1055 \\ \hline
No  $\pi^0$, no e-shower, No  protons                       &  53.6\% &  0    & 0   &  7830 &  2924 &  824 &  444 \\ \hline
0.8 GeV $<$ E$_{vis} <$ 1.05 GeV  &  49.8\% &  0    & 0   &  1064 &   408 &  137 &   56 \\ \hline
$p_{tot} <$ 0.35 GeV              &  44.8\% &  0    & 0   &    18 &     2 &    5 &    1 \\ \hline
\end{tabular}
\caption{\small Detailed list of cuts for the considered neutron decay
 channels. Survival fraction of signal (first column) and background
 events through event selections applied in succession. Backgrounds are normalized to an exposure of 
1 Mton$\times$year.}
\label{tab:14}
\end{center}
\end{table*}
}

\subsection{Cosmic muon-induced background estimation}
\label{subsec:cosmicbkg}

As described in Section~\ref{subsec:muonbkg}, the cosmogenic background
has been computed in three steps. In the third step, neutrons, neutral kaons and
lambda's entering the detector were generated according to their expected energy
spectrum.
We simulated 2$\times 10^5$ events for each particle species and used values in
Table~\ref{tab:neutcomparison} for normalization. The neutral particles
are propagated inside the liquid Argon volume until they interact
inelastically or decay. The secondary particles produced in those processes
were used in the nucleon decay analysis.

As concrete examples, we discuss in detail the 
two proton decay channels, $p \rightarrow K^+   \; \bar{\nu}$ and 
$p \rightarrow \pi^+ \; \bar{\nu} $, and one neutron decay channel
$n \rightarrow \pi^0 \; \bar{\nu}$. Other decay channels with a lepton
in the final state or more
constrained kinematics and topologies, will be less affected
by cosmogenic backgrounds. For instance, the probability
to misidentify charged pions as muons is at these momenta
typically less than $10^{-2}$.

In the analysis, we started by applying loose kinematical cuts that rejected
cosmogenic events clearly incompatible with nucleon decay signals~\cite{thesisdai}. 
For instance, for the channel $p \rightarrow \pi^+ \; \bar{\nu}$,
the expected pion energy in an ideal detector would be 0.48~GeV.
We accepted charged pions 
with energy in the range 0.35--0.65~GeV,
since the measured energy is smeared by Fermi motion and detector effects.
In the case of strange mesons, we also required an identified 
kaon in the final state and a total energy below 0.8~GeV. 
In a second step, same cuts as those defined in Tables~\ref{tab:1} and
\ref{tab:14} were applied to estimate final background contaminations.

Table~\ref{tab:cosmubkg1} summarizes for each considered
channel the remaining muon-induced 
background level
after this selection. 
The contamination coming from neutrons, kaons and lambdas interactions
 at different detector depths are shown. 
 We found that the $\pi^\pm$ and $\pi^0$ backgrounds are
predominantly produced by neutrons, and the $K^\pm$ background is dominated
by the $K^0_L$ entering into the detector. For some channels we obtain zero
events after the final cut, hence an upper limit for these events occurring
in the detector is given. The $\Lambda$ induced background events appeared
to be negligible in comparison to the other sources.

\begin{table*}
\tabcolsep=.8mm
\centering
\begin{tabular}{|l|c|c|c|c|c|c|c|}
\hline
& &  \multicolumn{3}{c|}{Background source $N^0_b$} & \multicolumn{3}{c|}{Cosmogenic background reduction} \\
 Depth & Channel &
 \multicolumn{3}{c|}{(particles/year)}  & Distance cut & Fiducial mass & Background $N_b$ \\
 & &
 \multicolumn{1}{c|}{Neutron} & \multicolumn{1}{c|}{K$^0$} &
 \multicolumn{1}{c|}{$\Lambda$} & d (m) & (kton) & (events/year) \\
\hline\hline
    $\simeq $0.5~km w.e.         & $p\rightarrow \pi^+\bar\nu$ & 570    & -- & -- 
    & 1.5  &  92 &  76 \\
(188~m rock)  & $n\rightarrow \pi^0\bar\nu$     & 450    & -- &        8        
& 1.7  &  91 &  46   \\
      \FLUKA{}      & $p\rightarrow K^+\bar\nu$   & --  &         135           & --  
      & 6.6  &  66 &  0.1\\
\hline\hline
 $\simeq $  1~km w.e.          & $p\rightarrow \pi^+\bar\nu$ & 200    & --  & -- 
 & 0.7  &  96 & 77 \\
 (377~m rock) & $n\rightarrow \pi^0\bar\nu$     & 130    & --  &        2.3     
 & 0.75 &  96 & 47      \\
      \FLUKA{}      & $p\rightarrow K^+\bar\nu$   & --  &         39           & --  
      & 5.45  &  71 &  0.1\\
\hline\hline
$\simeq $  3~km w.e.           & $p\rightarrow \pi^+\bar\nu$ &       4.0            & -- & -- 
& 0    & 100 &  4.0\\
(1.13~km rock)  & $n\rightarrow \pi^0\bar\nu$     &       2.6            & -- & -- 
& 0    & 100 &  2.6\\
     \FLUKA{}       & $p\rightarrow K^+\bar\nu$   & --  &         0.74         & --    
     & 1.8  &  90 &  0.1\\
\hline
\hline
  Under the hill & $p\rightarrow \pi^+\bar\nu$ & 2900    & --  & --  
  & 2.7  &  85 &  76 \\
(see Figure~\ref{fig:geometry3d})   & $n\rightarrow \pi^0\bar\nu$     & 2300    & --  &     --       
& 2.9  &  84 &  46 \\
\GEANT{}    & $p\rightarrow K^+\bar\nu$   & --  &         36--360           & -- 
& 5.4--7.5  &  72--62 &  0.1 \\

\hline
 Under the hill  & $p\rightarrow \pi^+\bar\nu$ & 430   & -- & -- 
 & 1.3  &  93 &  76 \\
+ two veto planes & $n\rightarrow \pi^0\bar\nu$     & 340 & --  &     --    
 & 1.5  &  92 &  46    \\
 \GEANT{}    & $p\rightarrow K^+\bar\nu$   & --  &         5--54           & -- 
 & 3.65--5.75  &  80--70 &  0.1 \\

\hline
 Under the hill  & $p\rightarrow \pi^+\bar\nu$ & 170    & --  & -- 
 & 0.6  &  97 &  77\\
+ three veto planes & $n\rightarrow \pi^0\bar\nu$     & 140    & --  &     --        
& 0.8  &  95 &  46\\
\GEANT{}    & $p\rightarrow K^+\bar\nu$   & --  &         2--20          & --  
& 2.8--5  &  85--74 &  0.1\\
\hline
\end{tabular}
\caption{Cosmogenic background for three selected channels: estimated number of background events per year that survive a kinematic 
selection. The contamination coming from neutrons, kaons and lambdas interactions
 at different detector depths are shown. 
 For each detector depth, the radial cut distance and the final fiducial volume to reduce cosmogenic background
 to the level of the irreducible atmospheric background (resp. 78.2 for
 $p\rightarrow \pi^+\bar\nu$, 47.4 for  $n\rightarrow \pi^0\bar\nu$ and
 0.1 for  $p\rightarrow K^+\bar\nu$ for an exposure of 100~kton$\times$year) 
 is listed. The range for kaon background is reflecting
 uncertainty on kaon yields due to differences between \FLUKA{} and \GEANT{} results.}
\label{tab:cosmubkg1}
\end{table*}

\begin{figure}
\centering
\includegraphics[width=0.30\textwidth]{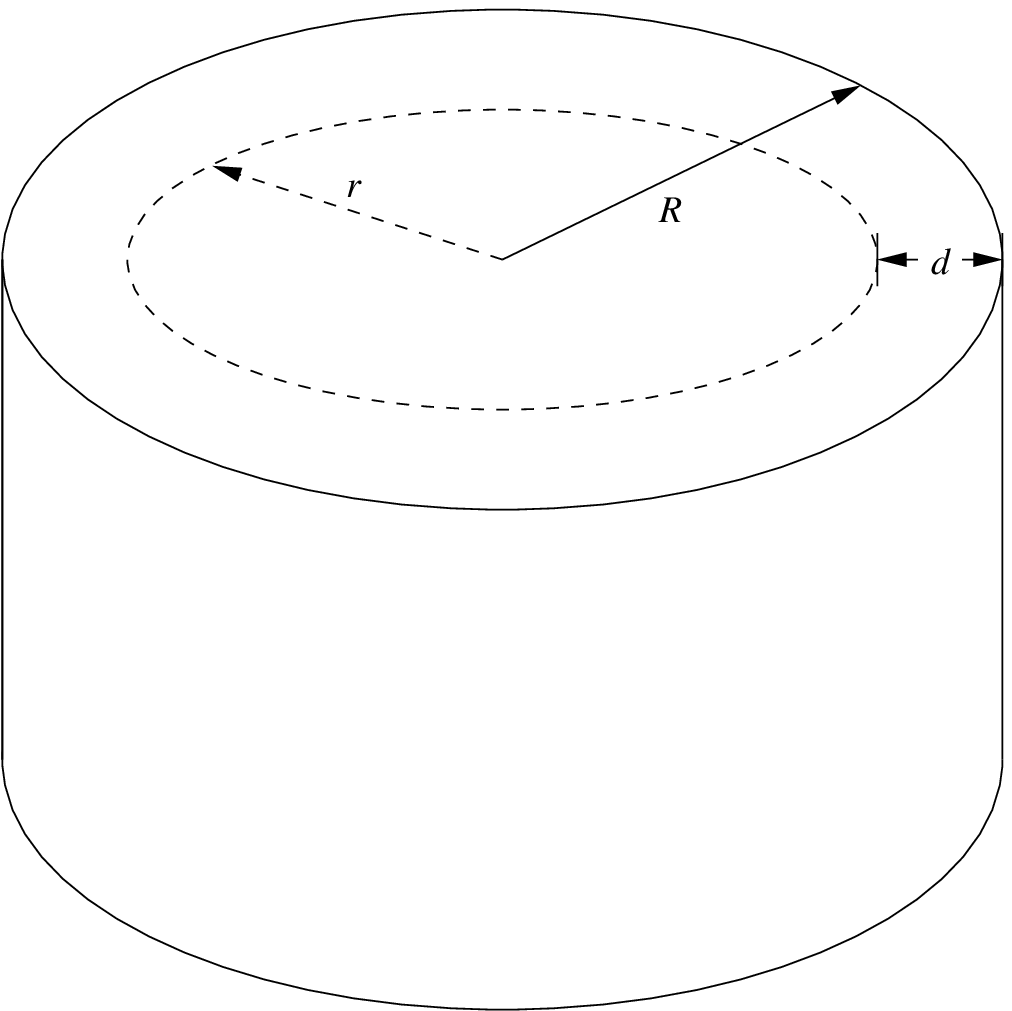} 
\caption{Schematic view of the radial distance cut to suppress cosmogenic
neutral particle background entering the sides of the detector.}
\label{Fig:dt1}
\end{figure}

The photo-nuclear
interactions, from which at least one neutral particle enters
the detector, occur mostly at an average distance of one meter
from the detector walls (see Figure~\ref{fig:rzeta}). 
Moreover, the number of background events reduces itself essentially exponentially
along their path inside the detector due to the self shielding properties
of Argon. Therefore, an obvious
action to reduce background
consists on cutting on the sides of the detector, excluding events produced at a
distance smaller than $d$ from the wall (see Figure~\ref{Fig:dt1}).
The fiducial volume is reduced accordingly. 
The value for the cut distance $d$ for each detector depth can be 
chosen in such a way that the remaining muon-induced background 
is of the same order than the irreducible atmospheric neutrino
background. 

In order to compute the effect of the distance cut on the background events,
we consider here the most conservative situation, i.e. the incident neutral background source
particles ($n$ and $K^0$) move horizontally from the top/bottom side wall into the detector (the shortest possible way).  After reaching the cut distance $d$, the particles move along the diagonal of the inner detector cylinder (the longest possible way). 
We obtain then a simplified
expression for the normalised background events:
\begin{equation} 
\label{eq:zc3}
N_b=N^0_b e^{-\frac{d}{c_\lambda}}\frac{R^2}{(R-d)^2}
\end{equation} 
where $N^0_b$ and $N_b$ are the numbers of events before and after the cut on the fiducial volume, and $c_\lambda$ is an effective interaction length found to be $\sim 71$~cm for neutrons and $\sim 86$~cm for kaons from the MC simulations.
With the given number of neutrino produced background events as $N_b$, the cut distance $d$ can be calculated from the following  equation:
\begin{equation} 
\label{eq:cd3}
d=R+2c_\lambda \text{W}\left(-\frac{R}{2c_\lambda}\sqrt{\frac{N^0_b}{N_b}} e^{-\frac{R}{2c_\lambda}}\right)
\end{equation}
where the $W(x)$ is the Lambert W-function (or Omega function,
inverse of $f(w)=w e^{w}$). 
%

Table~\ref{tab:cosmubkg1} summarizes, 
for each considered channel,
the distance cuts applied together with the remaining fiducial mass.
For each detector depth, the radial cut distance and the final fiducial volume to reduce cosmogenic background
 to the level of the irreducible atmospheric background (resp. 78.2 for
 $p\rightarrow \pi^+\bar\nu$, 47.4 for  $n\rightarrow \pi^0\bar\nu$ and
 0.1 for  $p\rightarrow K^+\bar\nu$ for an exposure of 100~kton$\times$year) 
 is listed. The range for kaon background is reflecting
 uncertainty on kaon yields due to differences between \FLUKA{} and \GEANT{} results.
 
We observe that the background due to charged and neutral pions can be
reduced to the same level of expected neutrino-induced backgrounds
without a big loss of the detector fiducial mass, even at shallow 
depth (1 km w.e. or equivalently 380 m of ordinary rock). The
rejection of strange mesons requires harder cuts. For a 1 km
w.e. depth, a reduction of 30$\%$ of the mass is expected. However the
prospects for a shallow depth experiment are promising and good
sensitivities for most of the analyzed channels 
are expected for an underground location of 1 km w.e. 
(see Table~\ref{tab:results}). Similar conclusions hold when a 0.5 
km w.e. depth (about 200 m of ordinary rock) is considered. 
Clearly the situation worsens very much 
if we consider a position close to surface. For a 50 m rock
overburden, the reduction of the pion background requires a reduction by 
more than $30\%$ of the usable Argon. The reduction of the kaon
background implies that half of the available Argon should be used
as shield. 

Note that using an annular veto system to detect muons as explained in Section~\ref{sec:RPC}, the induced background can be reduced by a factor between $\sim 5$ and $\sim 15$ (according to the number of planes used), therefore a looser cut on the distance $d$ would be
needed and it would be possible to have a bigger fiducial volume.
For example, the under the hill configuration with three veto planes consents 
to reduce the distance cut from $\approx 6$~m to $\approx 3$~m (see Table~\ref{tab:cosmubkg1}).


\section{Sensitivity to nucleon decay and comparison to Super-Kamiokande
results}
\label{sec:sensitivity}

We summarize here the the results from the previous sections for all channels
(see Table~\ref{tab:channels}) and compute partial lifetime sensitivities.

In case no signal is observed, limits to proton and neutron partial 
lifetimes $\tau/B$ will be obtained using : 
\begin{equation}
(\tau / B)_{proton} > \frac{2.7}{S} \times T \times \epsilon \times
10^{32} \; \; \; \; \mathrm{(years)} 
\end{equation}
\begin{equation}
(\tau / B)_{neutron} > \frac{3.3}{S} \times T \times \epsilon \times
10^{32} \; \; \; \; \mathrm{(years)} 
\end{equation}
T is the exposure in kilotons~$\times$ year,
$\epsilon$ the signal selection efficiency, and $S$ the constrained
90\% CL upper limit on the number of observed signal events,
taking into account the fact that there
are about 2.7 $\times 10^{32}$ protons and 3.3 $\times 10^{32}$
neutrons in 1 kton of Argon. $S$ is found by solving the equation~\cite{Yao:2006px}:
\begin{equation}
\frac{\sum_{n=0}^{n_0} P(n,b+S)}{\sum_{n=0}^{n_0} P(n,b)} = \alpha 
\end{equation}
$P(n,\mu)$ is the Poisson function,
$b$ the estimated background, 
$\alpha = 0.1$ for a~90\%~CL, and,
$n_0$ is equal to the closest integer number to~$b$ when computing
the ``detector sensitivity''.

\begin{table*}
\centering
 \begin{tabular}{|l|c|c|c|c|}
\hline
\multicolumn{1}{|c|}{Channel} & Cut & Total background &   $\tau$/B limit  (years) & $\tau$/B limit  (years) \\
                              &  efficiency (\%)               & per year  in fiducial volume    &       1 year exposure    &  10 years exposure      \\
\hline\hline
(p1) $p \rightarrow e^+ \; \pi^0$            & 45.3 &  0.1   & 0.5 $\times 10^{34}$ &   0.4 $\times 10^{35}$ \\
\hline
(p2) $p \rightarrow \pi^+ \; \bar{\nu} $     & 41.9 &  82 (3~km w.e.)  & 0.7 $\times 10^{33}$& 0.3 $\times 10^{34}$ \\
                                        &      &  151 (1~km w.e.)& 0.5 $\times 10^{33}$& 0.2 $\times 10^{34}$\\
                                        &      &  148 (0.5~km w.e.)& 0.5 $\times 10^{33}$& 0.2 $\times 10^{34}$\\
                                       &      &  143 (Under the hill) & 0.4 $\times 10^{33}$& 0.2 $\times 10^{34}$\\
                                        &      &  149 (Under the hill+2 veto planes)& 0.5 $\times 10^{33}$& 0.2 $\times 10^{34}$\\
                                       &      & 152 (Under the hill+3 veto planes)& 0.5 $\times 10^{33}$& 0.2 $\times 10^{34}$\\
\hline
(p3) $p \rightarrow K^+   \; \bar{\nu}$      & 96.8 & 0.2 (3~km w.e.)  & 1.0 $\times 10^{34}$ &   0.6 $\times 10^{35}$ \\
                                        &      & 0.2 (1~km w.e.)  & 0.8 $\times 10^{34}$&   0.6 $\times 10^{35}$\\
                                         &      &  0.2 (0.5~km w.e.) & 0.8 $\times 10^{34}$&   0.4 $\times 10^{35}$\\
                                         &      & 0.2 (Under the hill) & (0.8--0.7) $\times 10^{34}$&   (0.5--0.4) $\times 10^{35}$\\
                                         &      & 0.2 (Under the hill+ 2 veto planes)& (0.9--0.8) $\times 10^{34}$&   (0.5--0.5) $\times 10^{35}$\\
                                         &      &  0.2 (Under the hill+ 3 veto planes) & (1.0--0.8) $\times 10^{34}$&   (0.6--0.5) $\times 10^{35}$\\
\hline
(p4) $p \rightarrow \mu^+ \;   \pi^0   $     & 44.8 &  0.8   & 0.4 $\times 10^{34}$ &   0.2 $\times 10^{35}$\\
\hline
(p5) $p \rightarrow \mu^+ \; K^0 $           & 46.7 &  $<0.2$   & 0.5 $\times 10^{34}$  &   0.5 $\times 10^{35}$\\
\hline
(p6) $p \rightarrow e^+ \; K^0 $             & 47.0 &  $<0.2$   & 0.5 $\times 10^{34}$ &   0.5 $\times 10^{35}$\\
\hline
(p7) $p \rightarrow e^+ \; \gamma $          & 98.0 &  $<0.2$   & 1.1 $\times 10^{34}$ &   1.1 $\times 10^{35}$\\
\hline
(p8) $p \rightarrow \mu^+ \; \gamma $        & 98.0 &  $<0.2$   & 1.1 $\times 10^{34}$ &   1.1 $\times 10^{35}$\\
\hline
(p9) $p \rightarrow \mu^- \; \pi^+ \; K^+$   & 97.6 &  0.1   & 1.1 $\times 10^{34}$  &   0.8 $\times 10^{35}$\\
\hline
(p10) $p \rightarrow e^+ \; \pi^+ \; \pi^- $  & 18.6 &  2.5   & 0.1 $\times 10^{34}$  &   0.5 $\times 10^{34}$\\
\hline\hline
(n1) $n \rightarrow \pi^0 \; \bar{\nu}$ & 45.1 & 50 (3~km w.e.)  & 0.1 $\times 10^{34}$ & 0.5 $\times 10^{34}$\\
                                   &      & 92 (1~km w.e.) & 0.1 $\times 10^{34}$ & 0.4 $\times 10^{34}$\\
                                   &      & 89 (0.5~km w.e.) & 0.1  $\times 10^{34}$& 0.4 $\times 10^{34}$\\
                                    &      & 86 (Under the hill) & 0.1 $\times 10^{34}$& 0.3 $\times 10^{34}$\\
                                     &      & 90 (Under the hill+ 2 veto planes)& 0.1 $\times 10^{34}$& 0.4 $\times 10^{34}$\\
                                      &      &  91 (Under the hill+ 3 veto planes)& 0.1 $\times 10^{34}$& 0.4 $\times 10^{34}$\\
\hline
(n2) $n \rightarrow e^-   \; K^+$       & 96.0 &   $<0.2$ & 1.4 $\times 10^{34}$ &   1.4 $\times 10^{35}$\\
\hline
(n3) $n \rightarrow e^+   \;   \pi^-$   & 44.4 &   0.8 & 0.4 $\times 10^{34}$&   0.2 $\times 10^{35}$\\
\hline
(n4) $n \rightarrow \mu^- \;   \pi^+$   & 44.8 &   2.6 & 0.4  $\times 10^{34}$&   0.2 $\times 10^{35}$\\
\hline
\end{tabular}
\caption{Summary table of the main nucleon decay search
results. 
For each channel, the signal detection efficiency and the total expected background 
are given together with the lifetime limit at 90\% CL for 1 and 10~years
of data-taking. For the $p\rightarrow K^+\bar\nu$
the range reflects the difference of yields of kaons per muon from \FLUKA{} and \GEANT{}.}
\label{tab:results}
\end{table*}


For each nucleon decay channel we have computed the $(\tau / B)$ limits
as a function of the exposure. This is done by rescaling the number of
expected background events at each exposure, and computing the corresponding
upper limit ($S$). The result, together with the detection signal efficiency
($\epsilon$), the total expected background and the final detector
mass (after fiducial cuts) are reported in Table~\ref{tab:results}. 
In the channels $p\rightarrow \pi^+\bar\nu$,
$p\rightarrow K^+\bar\nu$, and $n\rightarrow \pi^0\bar\nu$, the results at different 
detector depths are given separately.

\begin{figure*}
\centering
 \epsfysize=8.0cm\epsfxsize=8.0cm
 \hspace*{0.1cm}\epsffile{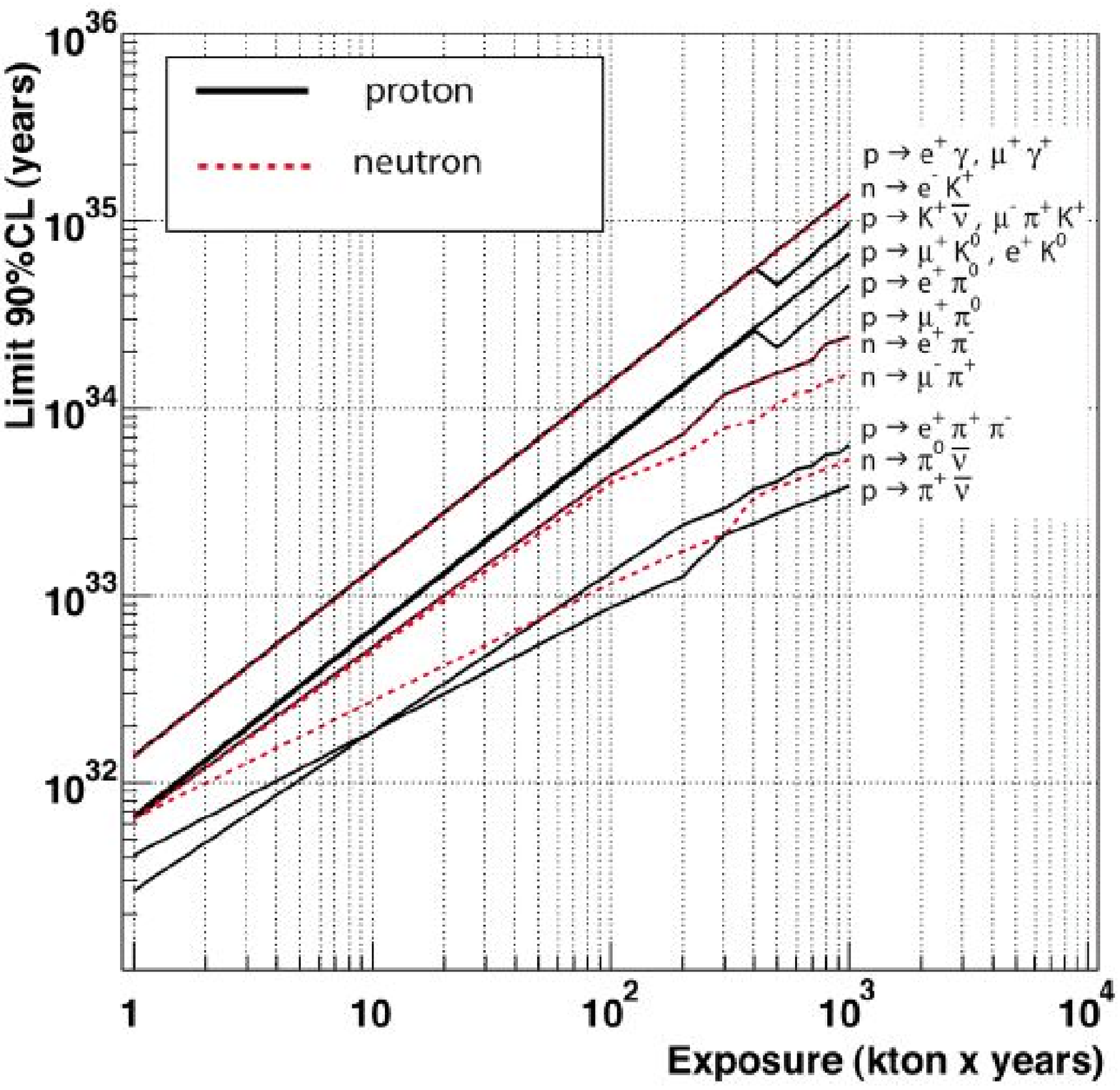}
 \epsfysize=8.0cm\epsfxsize=8.0cm
 \hspace*{0.1cm}\epsffile{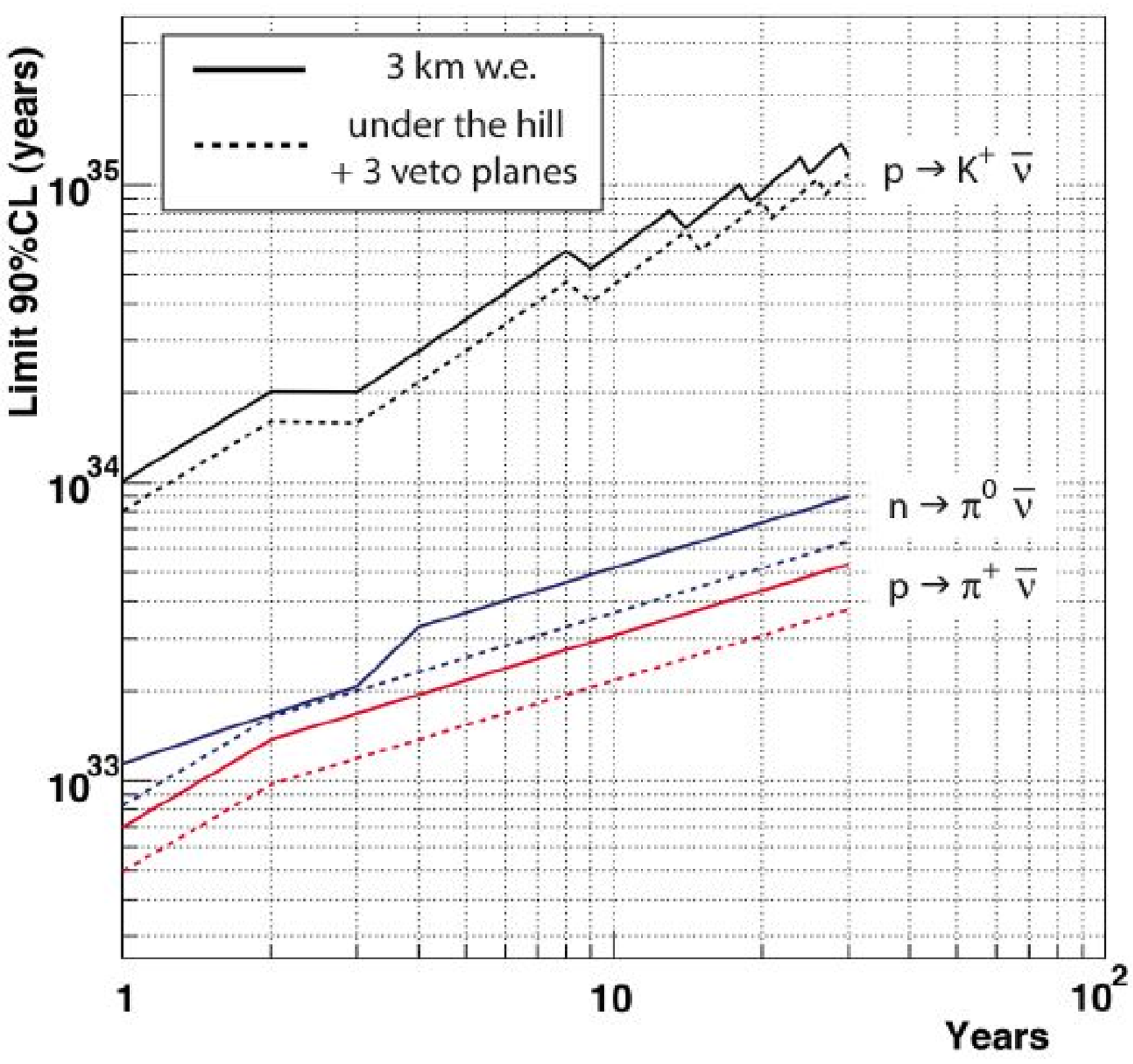}
 \caption{Running of the nucleon partial lifetime sensitivity ($\tau / B$ at 90\% C.L.)
 as a function of exposure (left) considering only atmospheric neutrino background 
 (right) with cosmogenic background and corresponding reduction 
 in fiducial mass taken into account (see text).}
 \label{fig:limit_pdk_expo}
\end{figure*}

The sensitivity of $(\tau / B)$ for protons and neutrons as a function of
the exposure is illustrated in Figure~\ref{fig:limit_pdk_expo} 
considering only atmospheric neutrino background (left) and
including the cosmogenic background (right). In the latter case 
the plot is shown as a function of exposure in years since there
is a reduction in fiducial mass taken into account to reduce
the background according to Table~\ref{tab:cosmubkg1}, while
the sensitivity curves without cosmogenic background are plotted
as a function of the exposure in kton~$\times$~year.

The liquid Argon TPC, offering good granularity and energy resolution, low particle detection threshold,
and excellent background discrimination, can  
yield good signal over background ratios in many possible
decay modes, allowing to reach partial lifetime sensitivities in
the range of $10^{34}-10^{35}$~years often in background
free conditions up to exposures of 1000~kton$\times$year.
This situation is optimal for discoveries at the few-events level.
In particular:
\begin{itemize}
\item Multi-prong decay modes like e.g. $p\rightarrow \mu^- \pi^+ K^+$
or $p\rightarrow e^+\pi^+\pi^-$ and channels involving kaons like
e.g. $p\rightarrow K^+\bar\nu$, $p\rightarrow e^+K^0$ and $p\rightarrow \mu^+K^0$
are particularly suitable, since liquid
Argon imaging
provides typically an order of magnitude improvement in efficiencies for similar
or better background conditions compared to Super-Kamiokande results.
\item Up to a factor 2 improvement in efficiency is expected for modes like $p\rightarrow e^+\gamma$
and $p\rightarrow \mu^+\gamma$ thanks to the clean photon identification
and separation from $\pi^0$. 
\item Channels like $p\rightarrow e^+\pi^0$ or $p\rightarrow \mu^+\pi^0$,
dominated by intrinsic nuclear effects,
yield similar efficiencies and backgrounds as in Super-Kamiokande. 
\end{itemize}

Thanks to the self-shielding and 3D-imaging properties of the liquid Argon TPC,
these results remains valid even at shallow depths where
cosmogenic background sources are important.
A very large area annular active muon veto shield could be used in order to
further suppress cosmogenic backgrounds at shallow depths.
For example, our results show that a three plane active veto at a shallow
depth of about 200~m rock overburden in the under the hill configuration yields
similar sensitivity for $p\rightarrow K^+\bar\nu$ as a 3~km~w.e. deep detector.

\section{Conclusions}
\label{sec.summary}

The most direct sign for Grand Unification is the experimental detection
of proton or bound-neutron decays. In order to reach partial lifetime
in the relevant range, new generation massive underground detectors 
with fine tracking and excellent calorimetry are needed to suppress 
backgrounds with a good signal selection efficiency.
Furthermore, the detector should be sensitive to several 
different channels in order
to better understand the nucleon decay mechanism. 

In this paper we have estimated the discovery potential 
of liquid Argon TPC detectors and compared it to the existing
results of Super-Kamiokande. We have analyzed 
many possible neutron and proton decay channels and for the first time
the two sources of background contamination: atmospheric neutrinos 
and cosmogenic backgrounds. 

Our results indicate that a liquid Argon TPC, if scaled at the relevant 
mass scale of the order of 100~kton,
offering good granularity and energy resolution, low particle detection threshold,
and excellent background discrimination, can  
yield good signal over background ratios in many possible
decay modes, allowing to reach partial lifetime sensitivities in
the range of $10^{34}-10^{35}$~years with exposures up to 1000~kton$\times$year,
 often in quasi-background-free conditions optimal for discoveries
 at the few events level, corresponding
to atmospheric neutrino background rejections of the order of $10^5$.

Multi-prong decay modes like e.g. $p\rightarrow \mu^- \pi^+ K^+$
or $p\rightarrow e^+\pi^+\pi^-$ and channels involving kaons like
e.g. $p\rightarrow K^+\bar\nu$, $p\rightarrow e^+K^0$ and $p\rightarrow \mu^+K^0$
are particularly suitable, since liquid
Argon imaging
provides typically an order of magnitude improvement in efficiencies for similar
or better background conditions compared to Super-Kamiokande.

Up to a factor 2 improvement in efficiency is expected for modes like $p\rightarrow e^+\gamma$
and $p\rightarrow \mu^+\gamma$ thanks to the clean photon identification
and separation from $\pi^0$. 

Channels like $p\rightarrow e^+\pi^0$ or $p\rightarrow \mu^+\pi^0$,
dominated by intrinsic nuclear effects,
yield similar efficiencies and backgrounds as in Super-Kamiokande. 

Thanks to the self-shielding and 3-D imaging properties of the liquid Argon TPC,
this result remains valid even at shallow depths where
cosmogenic background sources are important.
We consider the possibility of a very large area annular active muon veto shield in order to
further suppress cosmogenic backgrounds at shallow depths.

In conclusion, we find that
this class of detectors does not necessarily require very deep underground laboratories, like those typically
encountered in existing or planned sites, to perform very sensitive nucleon decay searches. 
As a concrete example, we considered a three-plane annular active veto surrounding a detector at a shallow
depth of about 200~m rock overburden in an under the hill configuration and found that
it yields similar sensitivities, in particular for $p\rightarrow K^+\bar\nu$,
as for a 3~km~w.e. deep detector.

In addition to a successful completion of the required technological R\&D necessary to reach 
the relevant mass scale of 100~kton in a cost-effective way, we point out the
importance of an experimental verification of the liquid Argon TPC physics potentialities
to detect, reconstruct and classify events in the relevant energy range. 
This experimental verification will require in addition
to possible specific tests in charged particle beams, the
collection and study of neutrino events in GeV range with statistics of 
the order of 100'000 events
or more.

\section{Acknowledgments}
\label{sec.acknowledgments}
This work was in part supported by ETH and the Swiss National  Foundation.
AB, AJM and SN have been supported by CICYT Grants FPA-2002-01835 
and FPA-2005-07605-C02-01. SN acknowledges support from the 
Ramon y Cajal Programme. We thank P.~Sala for help 
with \FLUKA{} while she was an ETH employee. We thank
A.~Marchionni for a careful reading of the manuscript.



\end{document}